\documentclass[twocolumn]{aastex63}
\usepackage{amsmath}
\usepackage{amssymb}
\usepackage{xcolor}
\usepackage{xspace}
\usepackage{graphicx}
\usepackage{mathtools}
\usepackage{courier}
\usepackage[T1]{fontenc}
\usepackage[normalem]{ulem}

\newcommand{\Ha}{H$\alpha$\xspace}

\newcommand{\NII}{[\ion{N}{2}]\xspace}
\newcommand{\FeII}{\ion{Fe}{2}\xspace}
\newcommand{\MgII}{\ion{Mg}{2}\xspace}
\newcommand{\MgIIDoubletWave}{\ion{Mg}{2} $\lambda\lambda2796, 2804$\xspace}
\newcommand{\MgIIDoubletLowWave}{\ion{Mg}{2} $\lambda2796$\xspace}
\newcommand{\MgIIDoubletHighWave}{\ion{Mg}{2} $\lambda2804$\xspace}
\newcommand{\OII}{[\ion{O}{2}]\xspace}
\newcommand{\OIIwave}{[\ion{O}{2}] $\lambda\lambda3727, 3729$\xspace}
\newcommand{\OIII}{[\ion{O}{3}]\xspace}

\newcommand{\Msun}{M$_\sun$\xspace}
\newcommand{\Mstar}{M$_*$\xspace}
\newcommand{\MsunPerYr}{\Msun yr$^{-1}$\xspace}
\newcommand{\YrInverse}{yr$^{-1}$\xspace}
\newcommand{\MsunPerYrPerSqrKpc}{\Msun\YrInverse kpc$^{-2}$\xspace}
\newcommand{\SigmaSFR}{$\Sigma_{\text{SFR}}$\xspace}
\newcommand{\vmax}{$v_{\text{max}}$\xspace}
\newcommand{\kms}{km s$^{-1}$}

\shorttitle{Connecting Galactic Outflows and Star Formation}
\shortauthors{Prusinski, Erb \& Martin}

\turnoffeditone

\begin{document}
\title{Connecting Galactic Outflows and Star Formation:\\ Inferences from H\boldmath{$\alpha$} Maps and Absorption Line Spectroscopy at \boldmath$1\lesssim z\lesssim 1.5$\unboldmath 
	\footnote{This work is based in part on observations taken by the 3D-HST Treasury Program (GO 12177 and 12328) with the NASA/ESA \textit{Hubble Space Telescope}, which is operated by the Association of Universities for Research in Astronomy, Inc., under NASA contract NAS5-26555.}
	\footnote{Some of the data presented herein were obtained at the W. M. Keck Observatory, which is operated as a scientific partnership among the California Institute of Technology, the University of California and the National Aeronautics and Space Administration. The Observatory was made possible by the generous financial support of the W. M. Keck Foundation.}}
\author[0000-0001-5847-7934]{Nikolaus Z. Prusinski}
\affiliation{The Leonard E.\ Parker Center for Gravitation, Cosmology and Astrophysics, Department of Physics, University of Wisconsin-Milwaukee, 3135 N Maryland Avenue, Milwaukee, WI, 53211, USA}
\affiliation{Cahill Center for Astronomy and Astrophysics, California Institute of Technology, MC 249-17, Pasadena, CA 91125, USA}	
\author[0000-0001-9714-2758]{Dawn K. Erb}
\affiliation{The Leonard E.\ Parker Center for Gravitation, Cosmology and Astrophysics, Department of Physics, University of Wisconsin-Milwaukee, 
3135 N Maryland Avenue, Milwaukee, WI, 53211, USA}
\author[0000-0001-9189-7818]{Crystal L. Martin}
\affiliation{Department of Physics, University of California Santa Barbara, Santa Barbara, CA 93106, USA}

\date{\today}	
\submitjournal{AJ}

\email{nik@astro.caltech.edu\\erbd@uwm.edu\\cmartin@physics.ucsb.edu}

\begin{abstract}

We investigate the connection between galactic outflows and star formation using two independent data sets covering a sample of 22 galaxies between $1 \lesssim z \lesssim 1.5$. The \textit{HST} WFC3/G141 grism provides low spectral resolution, high spatial resolution spectroscopy yielding \Ha emission line maps from which we measure the spatial extent and strength of star formation. In the rest-frame near-UV, Keck/DEIMOS observes \FeII and \MgII interstellar absorption lines, which provide constraints on the intensity and velocity of the outflows. We compare outflow properties from individual and composite spectra with the star formation rate (SFR) and SFR surface density (\SigmaSFR), as well as the stellar mass and specific star formation rate (sSFR). The \FeII and \MgII equivalent widths (EWs) increase with both SFR and \SigmaSFR at $\gtrsim 3\sigma$ significance, while the composite spectra show larger \FeII EWs and outflow velocities in galaxies with higher SFR, \SigmaSFR, and sSFR. Absorption line profiles of the composite spectra further indicate that the differences between subsamples are driven by outflows rather than the ISM. While these results are consistent with those of previous studies, the use of \Ha images makes them the most direct test of the relationship between star formation and outflows at $z>1$ to date. Future facilities such as the \textit{James Webb Space Telescope} and the upcoming Extremely Large Telescopes will extend these direct, \Ha-based studies to lower masses and star formation rates, probing galactic feedback across orders of magnitude in galaxy properties and augmenting the correlations we find here.

\end{abstract}


\section{Introduction}

Galaxy evolution is driven through the baryon cycle:\ cool gas flows into the galaxy from the cosmic web and is converted into stars, the most massive of which quickly die, expelling their metal-enhanced baryons into the interstellar medium \added{\citep{peroux_cosmic_2020}}. These ejecta can be propelled to the circumgalactic or intergalactic media through galactic winds.
Although the general outline is clear, a detailed understanding of the processes involved remains an open issue in modern astrophysics \added{\citep{veilleux_cool_2020}}. In this study, we focus on the relationship between galactic winds and star formation. 

Galaxies with intense star formation are observed to have powerful outflows of gas; however, the primary driving mechanisms remain uncertain. Active galactic nuclei (AGN) \citep{faucher-giguere_physics_2012}, energy injection from supernovae (SNe) \citep{leitherer_starburst99_1999,veilleux_galactic_2005}, cosmic rays \citep{grenier_nine_2015}, and radiation pressure \citep{murray_radiation_2011} generate and sustain galactic outflows\deleted{(see \citet{veilleux_cool_2020} for a recent review)}. At a given time, many of these mechanisms are occurring simultaneously and they are likely to interact in complex and nonlinear ways that depend on the type of galaxy \citep{hopkins_stellar_2012}. These outflows transfer energy and momentum from the centers of galaxies to large radii \citep{bouche_dynamical_2007,menard_probing_2011,chevalier_wind_1985,murray_maximum_2005}. In doing so, the outflows may cause a depletion in the availability of cool gas and star formation may be quenched \citep{tremonti_discovery_2007,hopkins_cosmological_2008,gabor_quenching_2011}.

The global star formation rate (SFR) reached its peak at $z\sim 2$ and since then, has been steadily decreasing \citep{hopkins_normalization_2006, bouwens_uv_2007}. A decrease in the rate of cool gas accretion onto galaxies may explain this drop in star formation \citep{keres_how_2005}, although a more robust understanding of galactic feedback is required to constrain the physical processes driving these changes. Simulations of feedback (e.g.\ \citealt{nelson_impact_2015,genel_galactic_2015,sales_feedback_2010}) and how it regulates galactic disk formation (e.g.\ \citealt{brooks_role_2009,minchev_formation_2015,sales_origin_2012,ubler_why_2014}) together with observations of gas flows at $1<z<2$ provide insight into galaxy and baryon cycle evolution during this critical period.

Galactic winds at $z\sim 1$ are typically traced by rest-frame UV absorption lines backlit by the stellar continuum \citep{weiner_ubiquitous_2009,rubin_persistence_2010,prochaska_simple_2011,erb_galactic_2012,kornei_properties_2012,martin_demographics_2012,bordoloi_dependence_2014}. Cool ($T\sim 10^4$ K) outflowing gas appears blueshifted with respect to the systemic velocity of the galaxy. These outflows are commonplace in starburst galaxies across cosmic time, from $z\sim 0$ \citep{heckman_nature_1990, strickland_starburst-driven_2000, martin_physical_2009,soto_emission-line_2012}, to $z\sim 1$ \citep{weiner_ubiquitous_2009,rubin_persistence_2010,zhu_near-ultraviolet_2015}, to $z\gtrsim 2$ \citep{steidel_structure_2010,shapley_physical_2011,schreiber_kmos_2019}. 

Observations and simulations have suggested a star formation rate surface density (\SigmaSFR) threshold needed to drive outflows of 0.1 \MsunPerYrPerSqrKpc, although many factors (e.g.\ galaxy escape velocity, inclination angle, etc.) contribute to the presence and detectability of outflows \citep{heckman_galactic_2002,murray_radiation_2011,kornei_properties_2012}.
Outflow velocities have been observed to increase with galaxy mass and star formation rate, suggesting that higher mass galaxies have more ambient gas and energy from SNe and radiation pressure (SNe rate and luminosity both scale with SFR) and therefore sustain faster outflows than their lower mass counterparts \citep{martin_mapping_2005, rupke_outflows_2005,chisholm_scaling_2015}. In addition, the equivalent widths of interstellar absorption lines associated with outflows are observed to increase with stellar mass, SFR, and SFR surface density \citep{weiner_ubiquitous_2009,rubin_persistence_2010,martin_demographics_2012,bordoloi_dependence_2014}.

In this paper, we observe star formation and outflow properties for a sample of galaxies at $1<z<1.5$. We probe interstellar \MgII and \FeII absorption with Keck/DEIMOS and use  WFC3/G141 grism data from the \textit{Hubble Space Telescope} to construct \Ha emission line maps that trace the spatial distribution of star formation for each object. We determine the equivalent widths of the absorption lines along with outflow velocities of the gas flows. Star formation rate surface densities for the same objects are ascertained through area measurements of the highest surface brightness regions of the \Ha maps. By combining the two data sets, we directly compare the structure of star formation to the outflow properties. 

The paper is organized as follows. We describe the joint dataset in Section \ref{sec:data}. Section \ref{sec:hst_data} shows how \Ha emission line maps were constructed, Section \ref{sec:sfr_sd} explains how star formation rate surface densities were calculated, and Section \ref{sec:keck_data} shows how outflow properties were measured. In Section \ref{sec:results}, we discuss the correlations between star formation and outflow properties. In Section \ref{sec:composite}, we form composite absorption line spectra, and in Section \ref{sec:summary}, we summarize and discuss our results. Throughout this paper, we use the \citet{salpeter_luminosity_1955} IMF and adopt the \citet{planck_collaboration_planck_2016} cosmology with $H_0=67.74$ km s$^{-1}$ Mpc$^{-1}$, $\Omega_m=0.3089$ and $\Omega_\Lambda=0.6911$. In this cosmology, at the median redshift of the sample ($z=1.22$), 1\arcsec\ corresponds to 8.5 kpc.

\section{Observations, Data Reduction, and Measurements}
\label{sec:data}

\begin{figure*}
	\centering
	\plotone{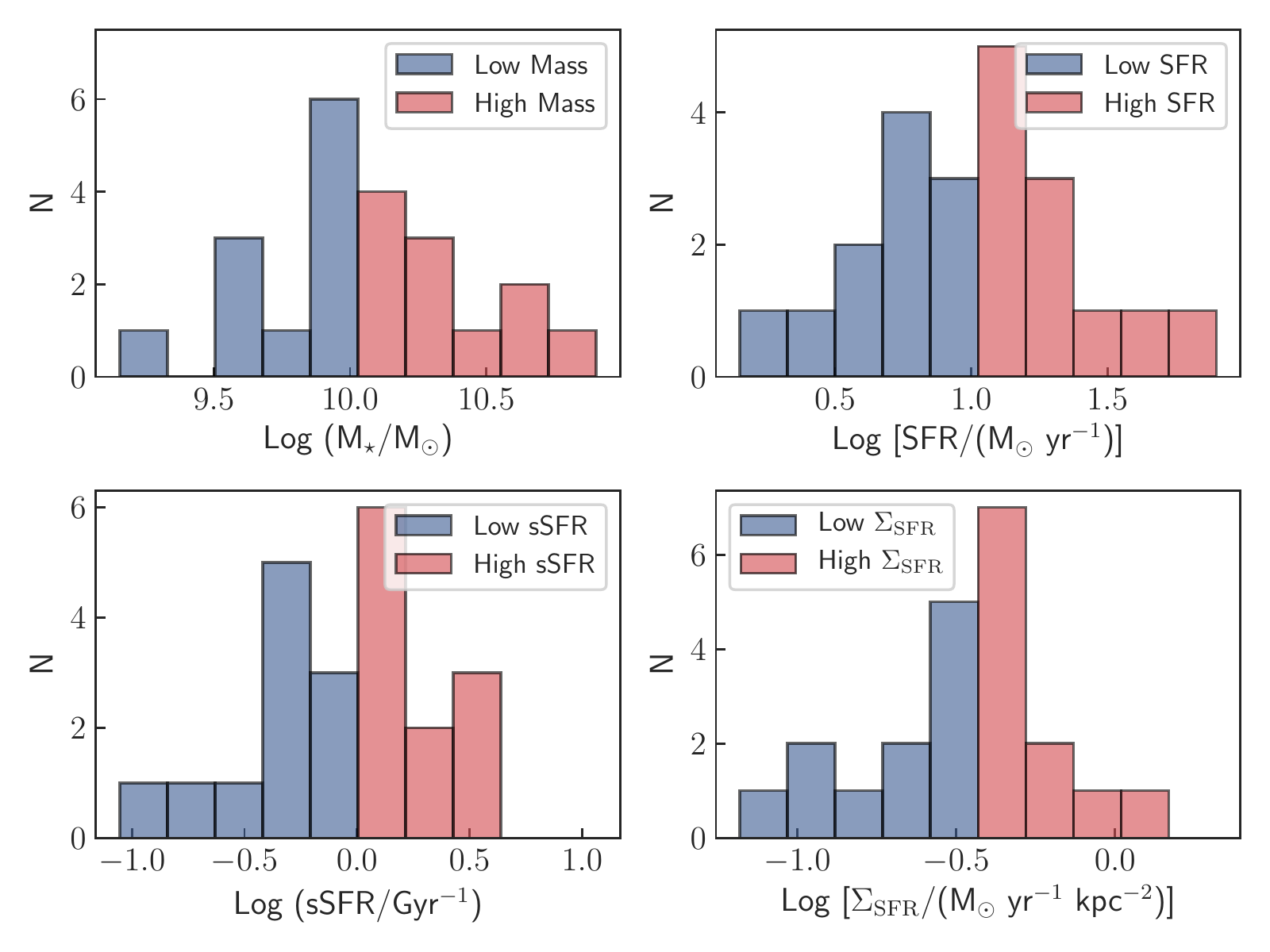}
	\caption{Distributions of galaxy properties for the 22 galaxy sample. Stellar mass (top left), star formation rate (top right), specific star formation rate (bottom left), and star formation rate surface density (bottom right) are shown. As described in Section \ref{sec:composite}, the full sample is split into two 11 object subsamples based on the median value of a given parameter. The blue and red histograms correspond to the low and high subsamples respectively from which composite spectra are formed.}
	\label{fig:hist}
\end{figure*}

In order to measure star formation and outflow velocities at $1\lesssim z \lesssim 1.5$, we constructed a joint data set comprising rest-frame near-UV absorption line spectra and rest-frame optical grism spectra to measure emission lines. Keck/DEIMOS detected \FeII and \MgII absorption lines which provided outflow velocities, and the \Ha emission line observed with \textit{HST} WFC3/G141 traces star formation.

The galaxies in this sample were selected from the Cosmic Assembly Near-IR Deep Extragalactic Legacy Survey (CANDELS), specifically the Extended Groth Strip (EGS; $\alpha=$ 14:19:31, $\delta=$ +52:51:00) and the Cosmic Evolution Survey (COSMOS; $\alpha=$ 10:00:29, $\delta=$ +02:20:36) fields \citep{grogin_candels_2011,koekemoer_candels_2011}. The sample was chosen using data from the \citet{skelton_3d-hst_2014} photometric catalog such that each galaxy had (1) a SFR $>$ 1 \MsunPerYr \added{as measured by spectral energy distribution (SED) fitting from the photometric catalog}; (2) $0.7\lesssim z_{\text{phot}}\lesssim 1.6$ at 99\% confidence so that the systemic redshift $z_{\text{sys}}$ could be measured from the \OIIwave doublet and absorption lines from \FeII ($\sim2300-2600$ \AA) and \MgII (2800 \AA) were visible; and (3) $\mathcal{R}\lesssim 24$ in order to obtain spectra with continuum S/N sufficient to measure absorption lines in individual objects in one night of observation. \added{Starting from the full COSMOS and EGS catalogs, we eliminated objects with unreliable photometry or poorly constrained SED fits, which when combined with the above three criteria reduced the 29,791 (36,699) objects in the COSMOS (EGS) field to a sample of 520 (446) galaxies. Of the three requirements, the magnitude cut eliminated the largest number of objects from the sample. Because our targets are bright, the catalog from which they are drawn is highly complete over our magnitude range of interest; \citet{skelton_3d-hst_2014} estimate a 90\% completeness level at $H_{\rm F160W}=25.1$, while the faintest object in our final sample is two magnitudes brighter than this with $H_{\rm F160W}=23.1$. We therefore expect our parent sample to be highly complete, but note that our selection criteria may eliminate dusty galaxies with SFR $>$ 1 \MsunPerYr but $\mathcal{R}> 24$.}

Ground-based observations \added{for 84} of these galaxies were conducted on 2015 March 26 and 27 using the DEep Imaging Multi-Object Spectrograph (DEIMOS) on the Keck II telescope. DEIMOS is a medium-resolution optical spectrometer with spectral coverage from 4000 \AA\ to 10,500 \AA\ \citep{faber_deimos_2003}. \edit1{Out of the 84 galaxy sample at $1<z<1.5$, 47 objects had significant absorption line detections. These lines include \FeII $\lambda2344$, $\lambda2374$, $\lambda2383$, $\lambda2587$, $\lambda2600$, and \MgIIDoubletWave, which trace the outflow velocities of interstellar gas.} DEIMOS observations were conducted using the 600 lines mm$^{-1}$ grating with one slitmask per field and 1\arcsec\ slits. The dispersion was 0.65 \AA\ pixel$^{-1}$ and the spectral resolution FWHM was 4.6 \AA\ or $\sim 180$ \kms\ as measured from the widths of night sky lines. Total exposure times for the EGS and COSMOS fields were 8.79 hours and 9.04 hours respectively. The airmass ranged between 1.05 and 1.31, and the average seeing was $\sim$ 0\farcs 6.

\edit1{We further restrict our study to objects with detections of \Ha emission in the 3D-HST grism survey \citep{momcheva_3d-hst_2016,van_dokkum_first_2011,brammer_3d-hst_2012}.} Each of the galaxies in the EGS and COSMOS fields has WFC3 F140W+G141 direct and grism observations from two visits with an average exposure time of $\sim 5600$ s. The F140W filter and G141 grism have overlapping wavelength coverage from $\sim 12000-16000$ \AA, with \Ha visibility from $z\sim 0.75$ to $z\sim 1.5$ (2.75 Gyr of cosmic time). \edit1{Excluding objects with strong contamination (shown in Figure \ref{fig:2d-spec} and discussed in Section \ref{sec:hst_data}) and retaining objects with significant \Ha emission and either \FeII or \MgII absorption line detections led to a final sample of 22 galaxies (14 in COSMOS and 8 in EGS).} Of those 22 objects, 18 have significant \FeII detections, 20 have significant \MgII detections, and 16 have detections of both \MgII and \FeII absorption features. The median redshift of the sample is $1.22$ with standard deviation $0.16$, \added{and the galaxies generally fall on the star forming main sequence for that redshift \citep{speagle_highly_2014}}.

Figure \ref{fig:hist} shows the range of galaxy physical parameters for the 22 object sample. We observe a median stellar mass (\Mstar) of $1.1\times 10^{10}$ \Msun, a median SFR of 10.6 \MsunPerYr, a median specific star formation rate (sSFR) of 1.0 G\YrInverse, and a median SFR surface density of 0.4 \MsunPerYrPerSqrKpc.

\subsection{Rest-frame Optical Data from HST}
\label{sec:hst_data}

\begin{figure*}
	\centering
	\plottwo{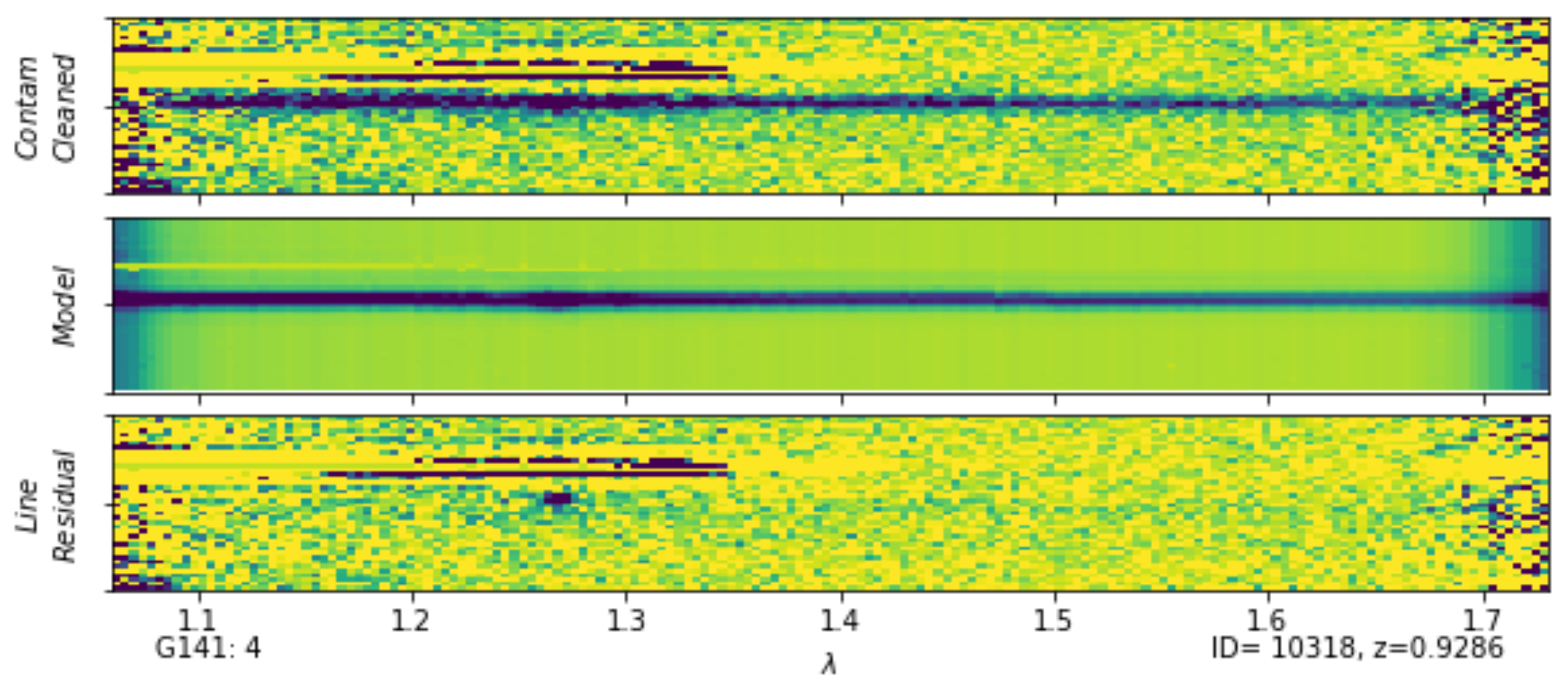}{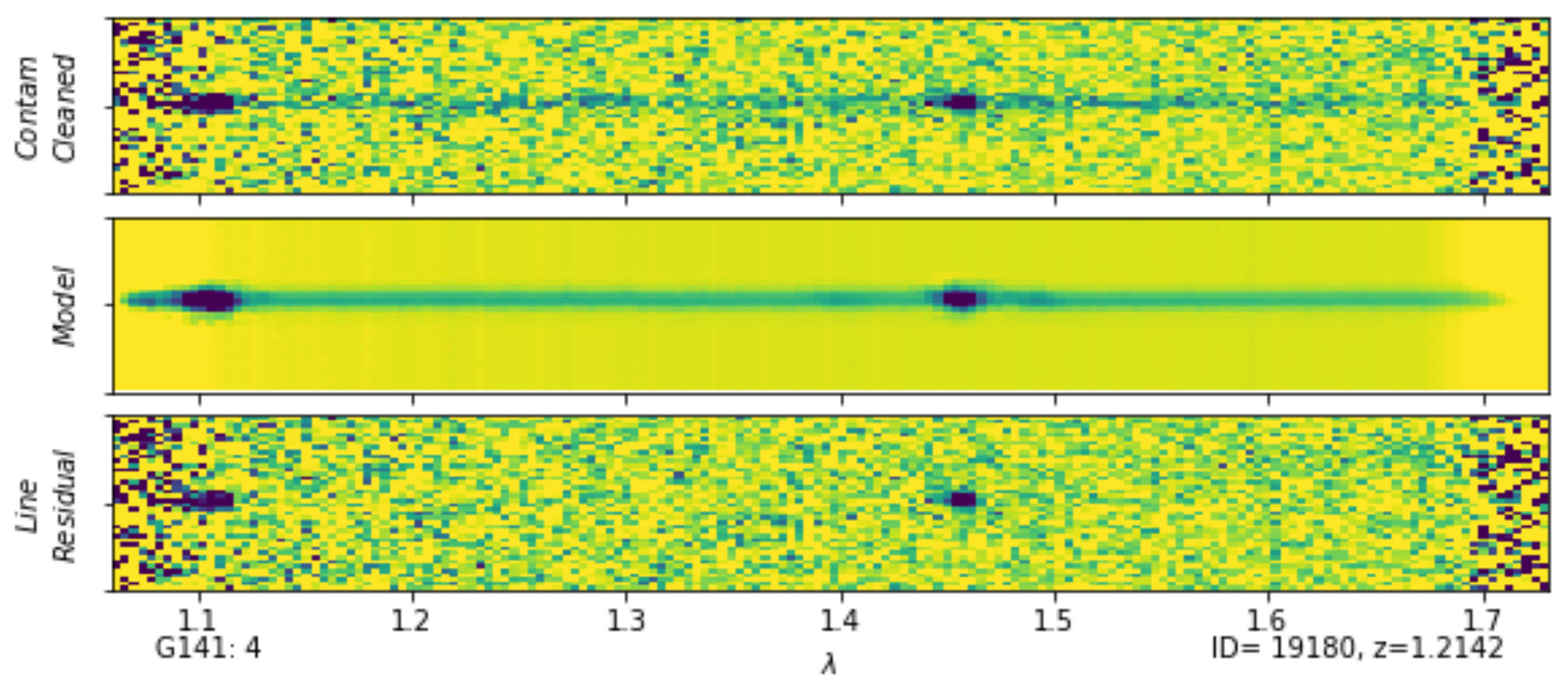}
	\caption{Examples of 2D spectra taken by the WFC3/G141 grism on \textit{HST}. The \textit{Grizli} pipeline produces a fully reduced 2D spectrum (top panel), model (middle panel), and subtracts the continuum to produce a line-only spectrum (bottom panel). On the left, COS 10318 has \Ha emission visible at 1.27 $\mu$m, while COS 19180 (right) has \OIII and \Ha emission lines present at 1.11 and 1.45 $\mu$m respectively. The dark bands in the spectrum of COS 10318 are contamination from another source in the field also visible in the \Ha map in Figure \ref{fig:cutouts}.}
	\label{fig:2d-spec}
\end{figure*}

The HST/WFC3 grism images were reduced using the Grism Redshift and Line Analysis (\textsc{Grizli}\footnote{\url{https://github.com/gbrammer/grizli/}}) pipeline (see \citealt{brammer_grizli_2019, abramson_grism_2020, wang_discovery_2019} for descriptions of \textsc{Grizli}). We input WFC3 direct images paired with G141 grism images containing dispersed 2D spectra for each of the objects in the field. In addition, we supply the spectroscopic redshift of the target galaxy given by the \OII emission line observed in the Keck/DEIMOS spectra. \textsc{Grizli} first pre-processes the G141 exposures by performing astrometric alignment, background subtraction, flat-fielding, and extracting visit-level catalogs and segmented images from the corresponding direct image. Using the \texttt{AstroDrizzle} software \citep{gonzaga_drizzlepac_2012}, the pipeline returns drizzled mosaics of the direct and grism images. \textsc{Grizli} then makes continuum and contamination models using a polynomial fit and extracts 1D and 2D spectra. Examples of the 2D spectra are shown in Figure \ref{fig:2d-spec}.

\begin{figure*}
	\plotone{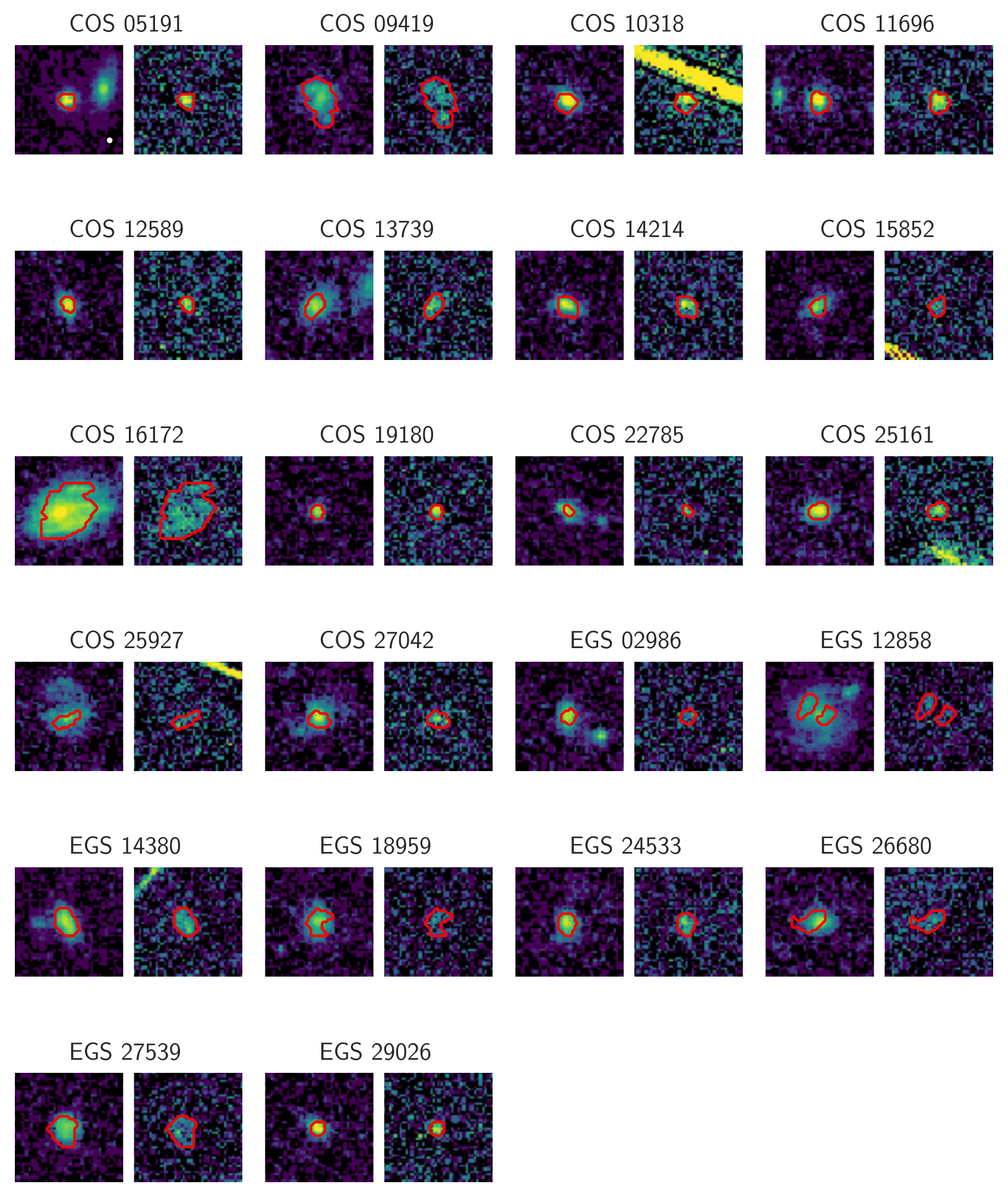}
	\caption{F140W direct images (left panel) and \Ha line maps (right panel) of the objects in the sample. Each image is 4\arcsec\ on a side with the red contour indicating the area maximizing \Ha S/N. The WFC3/G141 PSF has a FWHM of 0\farcs 136 and is denoted by the white circle in the top left plot. All plots are on a log color scale. The direct images have a scale ranging between $5\times 10^{-22}$ and $2\times 10^{-20}$ erg s$^{-1}$ cm$^{-2}$ \AA$^{-1}$ while the \Ha emission line maps have a scale between $5\times 10^{-19}$ and $1\times 10^{-17}$ erg s$^{-1}$ cm$^{-2}$.}
	\label{fig:cutouts}
\end{figure*}

The next step is the creation of \Ha emission line maps. These are possible because of WFC3's high spatial resolution (0.136\arcsec) and G141's low ($R\sim 130$) point-source spectral resolution. A G141 grism spectrum comprises high spatial resolution images placed in series on the WFC3 detector in 46 \AA\ increments. These exposures cover the wavelength range (1075-1700 nm) of G141 and are placed sequentially on the WFC3 detector. To make the emission line maps, a spectral template is fit to the data using the given spectroscopic redshift of the galaxy. The maps are constructed by subtracting the stellar continuum model from the 2D spectrum; the remaining flux comes from emission features \citep{nelson_where_2016}. The direct image is used to map the spatial distribution of the emission line from the 2D spectrum to an 80 $\times$ 80 pixel postage stamp. With the pixel scale of 0.1\arcsec, this corresponds to 8\arcsec\ $\times$ 8\arcsec\ or 68 $\times$ 68 kpc at $z = 1.22$. In Figure \ref{fig:cutouts}, we plot the central $4\arcsec\times 4\arcsec$ region of each \Ha emission line map. The emission line in these maps appears as an image of the galaxy taken at the wavelength of the line. Some of these images contain contamination from other spectra in the field \edit1{(see left panel of Figure \ref{fig:2d-spec})}. We retained objects for which we were able to exclude regions with strong contamination, so that the contamination did not significantly impact the flux from the \Ha line. 

We apply two corrections to the \Ha flux in these maps:\ an \NII correction and an extinction correction. Due to the low spectral resolution of the G141 grism, the \Ha and \NII $\lambda 6583$ lines are blended, so we observe the combined flux of the two lines. After converting the galaxy masses in the photometric catalog \citep{skelton_3d-hst_2014} to a \citet{salpeter_luminosity_1955} IMF, we estimate the \NII emission as a function of galaxy mass using the \NII/\Ha mass-metallicity relation from \citet{yabe_massmetallicity_2014}. We then subtract the estimated \NII contribution from the total observed flux. The median \NII contribution across the sample is $\sim 17\%$ of the total flux. For the extinction correction, we assume the \citet{calzetti_dust_2000} extinction law and use A$_V$ values from the photometric catalog to apply an extinction correction to each of the \Ha fluxes; the median \Ha flux correction factor is 1.52. 

\added{We note that the adoption of a single value for A$_V$ assumes that the extinction does not vary across the galaxy, which may not be the case (e.g.\ \citealt{wang_uvi_2017}); however, the fact that we focus only on the central regions with the highest surface brightness \Ha emission likely mitigates this effect. It is also probable that the value of A$_V$ determined from SED fitting of the spatially integrated stellar continuum is not the value appropriate to the strongest \Ha emission, since nebular emission is generally more attenuated than the continuum \citep{calzetti_dust_2000,reddy_mosdef_2015} and extinction gradients measured from the Balmer decrement tend to peak in the centers of galaxies \citep{nelson_spatially_2016}. Since our galaxy sample does not have H$\beta$ emission measured over the same regions as \Ha we are unable to quantify this effect, but the potential impact is that the extinction corrections are underestimated and therefore the SFRs and SFR surface densities may be higher than reported.}

\subsubsection{Calculation of Star Formation Rate Surface Densities}
\label{sec:sfr_sd}

With dust and \NII-corrected \Ha fluxes in hand, we now seek to measure the galaxy area and compute star formation rate surface densities. Many previous studies (e.g.\ \citealt{rubin_persistence_2010,bordoloi_dependence_2014,heckman_systematic_2015,heckman_implications_2016}) have used an area proportional to $\pi r^2$, where $r$ is the half-light radius measured from space-based rest-frame UV imaging. Others have made minor variations to this method such as using the galaxy's semimajor axis \citep{rubin_evidence_2014} as opposed to the half-light radius; however, because star formation often occurs in small, separated clumps, sizes measured in this way may overestimate the area of the regions most likely to produce outflows \citep{rubin_persistence_2010}.  

\citet{kornei_properties_2012} take a more refined approach to measuring the SFR surface density, targeting regions of a galaxy with active star formation. For the subset of their sample in the redshift range for which the \citet{kennicutt_star_1998} conversion between UV luminosity and star formation can be applied, they calculate a ``clump'' area by selecting pixels above a \SigmaSFR threshold of 0.1 \MsunPerYrPerSqrKpc, since these are most likely to contribute to driving outflows. They then parameterize this area measurement by calculating a scale factor between the ``clump'' area and the area calculated using $\pi r_P^2$, where $r_P$ is the Petrosian radius. They find that the median ``clump'' area is 74\% of the area corresponding to the Petrosian radius, and then systematically define the area of each object to be the brightest region of the galaxy containing 74\% of the flux within the Petrosian radius.

Because our study has the advantage of using \Ha images, we can define areas that directly measure the regions of strongest star formation by converting the \Ha luminosity of each pixel to SFR via the \citet{kennicutt_star_1998} relation. On the other hand, the \Ha emission line maps have much lower S/N than the broadband images used to measure sizes in previous studies, and the S/N of individual pixels with \SigmaSFR $\sim$ 0.1 \MsunPerYrPerSqrKpc is generally low ($\sim 1$). For several reasons, we opt not to simply measure the area corresponding to the pixels above a specified SFR surface density threshold as suggested by \citet{kornei_properties_2012}. First, although there is some observational and theoretical justification for a threshold of \SigmaSFR~$= 0.1$~\MsunPerYrPerSqrKpc, this value is not robustly determined and we therefore prefer not to impose a somewhat arbitrary threshold onto the data. The use of a constant threshold also does not take the varying noise properties of the images into account, and because the S/N of the pixels near the proposed threshold is low it also results in the inclusion of significant noise.

Instead of using pixels above a particular threshold, we adopt a technique motivated by aperture photometry and define an optimal \Ha aperture designed to maximize the S/N of the integrated flux. In the case of background-dominated observations, the radius that maximizes the S/N of the enclosed flux is directly related to the characteristic scale of the surface brightness profile; for an exponential profile, the optimal radius is 1.8$h$, where $h$ is the scale length, while for a Gaussian profile it is 1.6$\sigma$ or 0.67 FWHM \added{(see Appendix \ref{app:sn} for a derivation)}. 

\edit1{We iteratively determine this optimal region by smoothing the emission line and error images slightly, using a range of smoothing kernels (typically 1–2 pixels). For each smoothed image, we create a series of apertures of varying sizes by selecting the pixels falling above a range of S/N thresholds in the smoothed data, and then measure the enclosed flux and its uncertainty for each of these apertures on the original, unsmoothed images. Finally, we choose the aperture that maximizes the \Ha S/N in the original data. } As expected, this method tracks the highest surface brightness regions, but has the advantages of not imposing a semi-arbitrary SFR surface density threshold and still allowing fainter pixels to contribute. We find that the median aperture size resulting from this method matches that found by instead including all pixels above a threshold of \SigmaSFR $> 0.18$ \MsunPerYrPerSqrKpc, although not all pixels within our apertures are above this threshold. The optimal \Ha apertures are shown by red contours in Figure \ref{fig:cutouts}. For comparison, these apertures are 40--90\% of the galaxy sizes measured from the direct F140W images by the \textsc{Grizli} pipeline. \added{As a result, \Ha fluxes measured using the full-light apertures from the direct images are $\sim 1.4$ times larger than those of the maximal S/N method.}

Once the sizes of the \Ha regions are defined, we convert the \Ha luminosity within each aperture to the star formation rate of the galaxy using the \citet{kennicutt_star_1998} relation, and find the star formation rate surface density by dividing the SFR by the area of the aperture. \edit1{We note that there is a systematic uncertainty associated with the conversion of \Ha luminosity to SFR, which \citet{kennicutt_star_1998} estimated to be $\sim 30\%$ from the variation in other published models and calibrations at the time}. The resulting median \SigmaSFR for the 22 galaxy sample is 0.4 \MsunPerYrPerSqrKpc; this is higher than that of \citet{kornei_properties_2012} because the lower S/N \Ha\ images are less effective in tracing fainter star formation. However, our sample spans more than an order of magnitude in \SigmaSFR, and we also find that four objects have a SFR surface density less than 0.18 \MsunPerYrPerSqrKpc, and one (EGS 12858) has \SigmaSFR~$<0.1$~\MsunPerYrPerSqrKpc. 

\added{For comparison with the literature, we also calculate SFR surface densities using the methodology of \citet{bordoloi_dependence_2014}, who adopt SFRs from SED fitting and define the SFR surface density to be $\Sigma_{\rm{SFR,\, B}} = \mbox{SFR}/2\pi R_{1/2}^2$, where $R_{1/2}$ is the half-light radius. Using SFRs from the \citet{skelton_3d-hst_2014} photometric catalog and half-light radii from the measurements of \citet{van_der_wel_structural_2012}, we find that the median SFR surface densities corresponding to our optimal apertures are a factor of $\sim 1.8$ times larger than the values of $\Sigma_{\rm{SFR,\, B}}$ for the sample. We discuss how correlations between these SFR surface densities and our outflow-related quantities compare with the optimal aperture results in Section \ref{sec:results}.} 

 The specific star formation rate of each galaxy is computed by dividing the SFR by the stellar mass determined from the \edit1{SED} in the 3D-HST catalog \citep{skelton_3d-hst_2014}. \added{Uncertainties on the stellar mass and A$_V$ are provided by R.\ Skelton (private communication), which we propagate into our error calculations below.} To establish uncertainties on each of \added{our measured} parameters, 350 Monte Carlo simulations were run for each galaxy. Each pixel in the emission line map was perturbed by a random amount drawn from a Gaussian distribution of width equal to the uncertainty on that pixel, and the measurements of area and SFR described above were repeated on each perturbed \Ha image. The error for each quantity was then estimated from the width of the 68\% confidence interval in the resulting distributions. The resulting measurements and uncertainties are given in Table \ref{tab:HST}.

\subsection{Rest-frame Near-UV Keck Spectroscopy}
\label{sec:keck_data}

\begin{figure*}
	\begin{minipage}{0.48\textwidth}
		\centering
		\includegraphics[width=\linewidth]{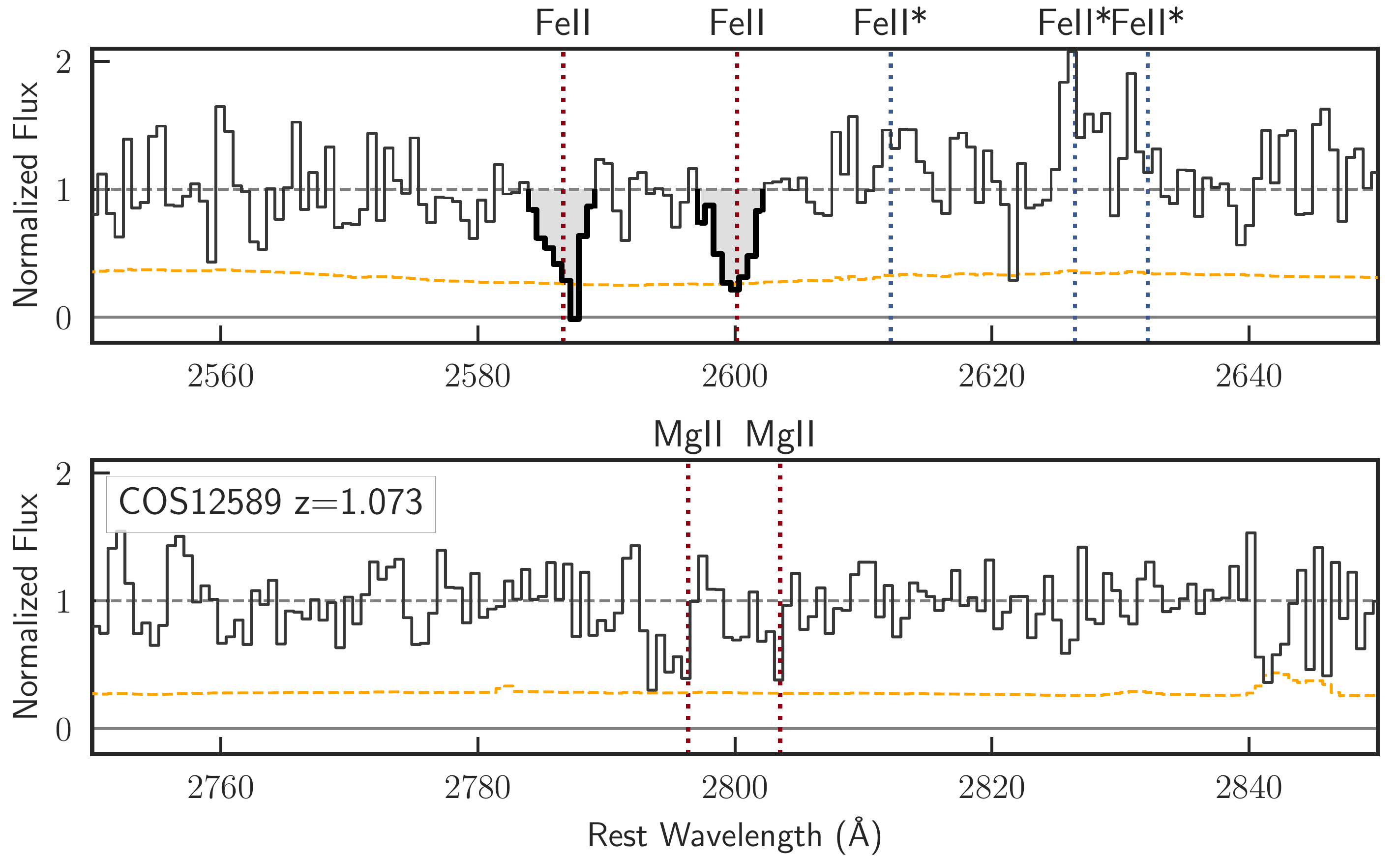}
	\end{minipage}
	\begin{minipage}{0.48\textwidth}
		\centering
		\includegraphics[width=\linewidth]{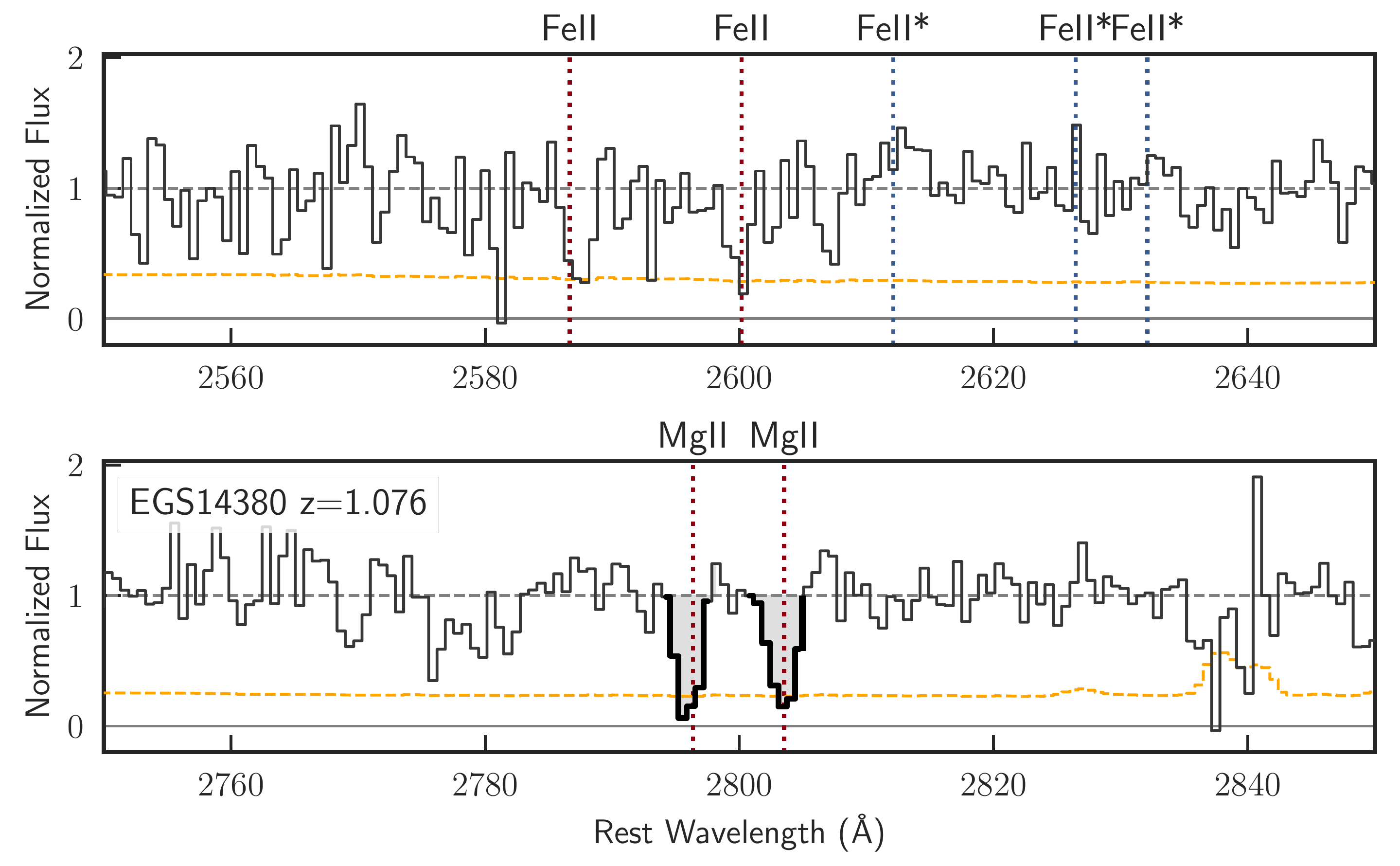}
	\end{minipage}
	\vspace{0.3cm}

	\begin{minipage}{0.48\textwidth}
		\centering
		\includegraphics[width=\linewidth]{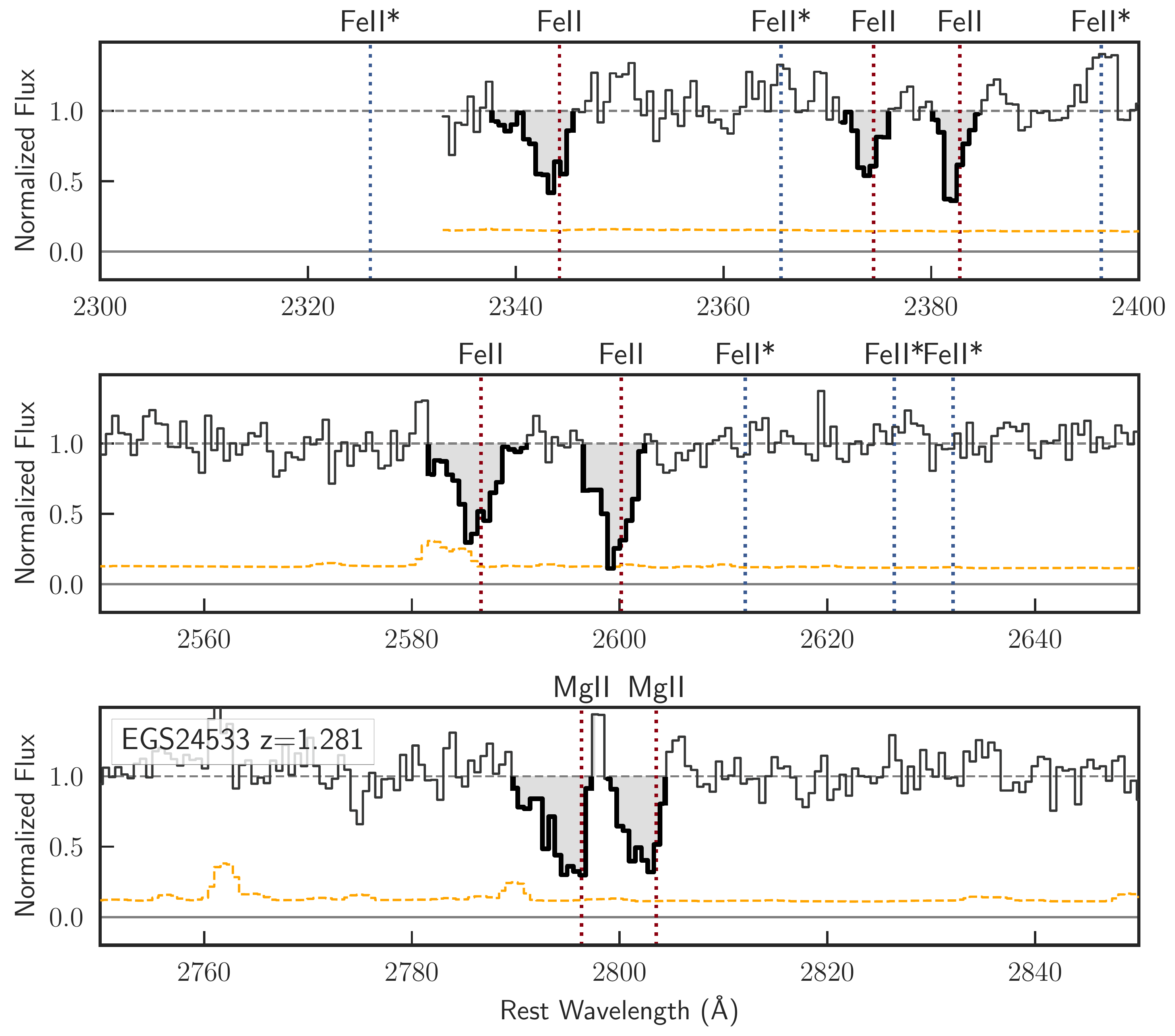}
	\end{minipage}
	\begin{minipage}{0.48\textwidth}
		\centering
		\includegraphics[width=\linewidth]{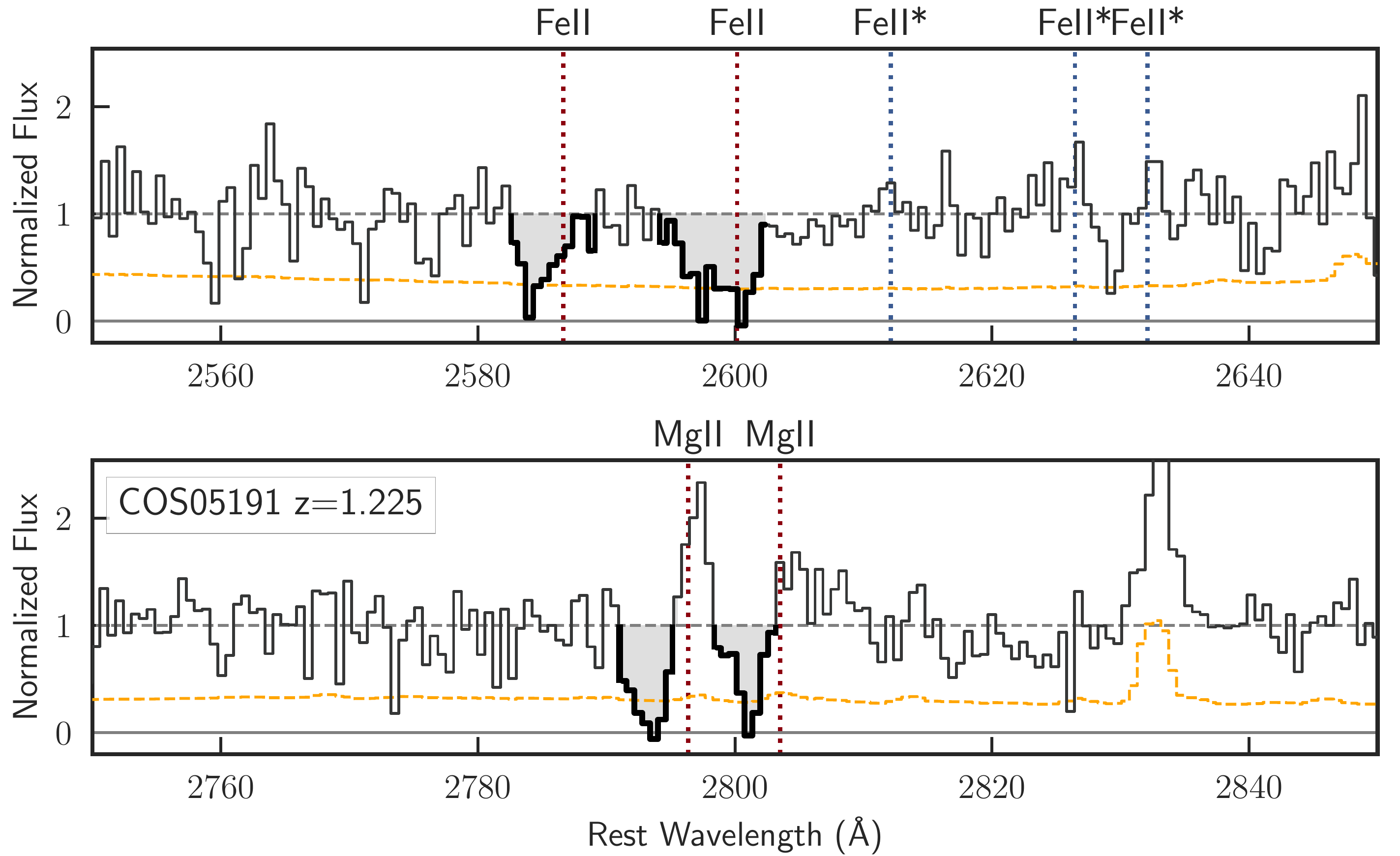}
	\end{minipage}
	\vspace{0.3cm}

	\begin{minipage}{0.48\textwidth}
		\centering
		\includegraphics[width=\linewidth]{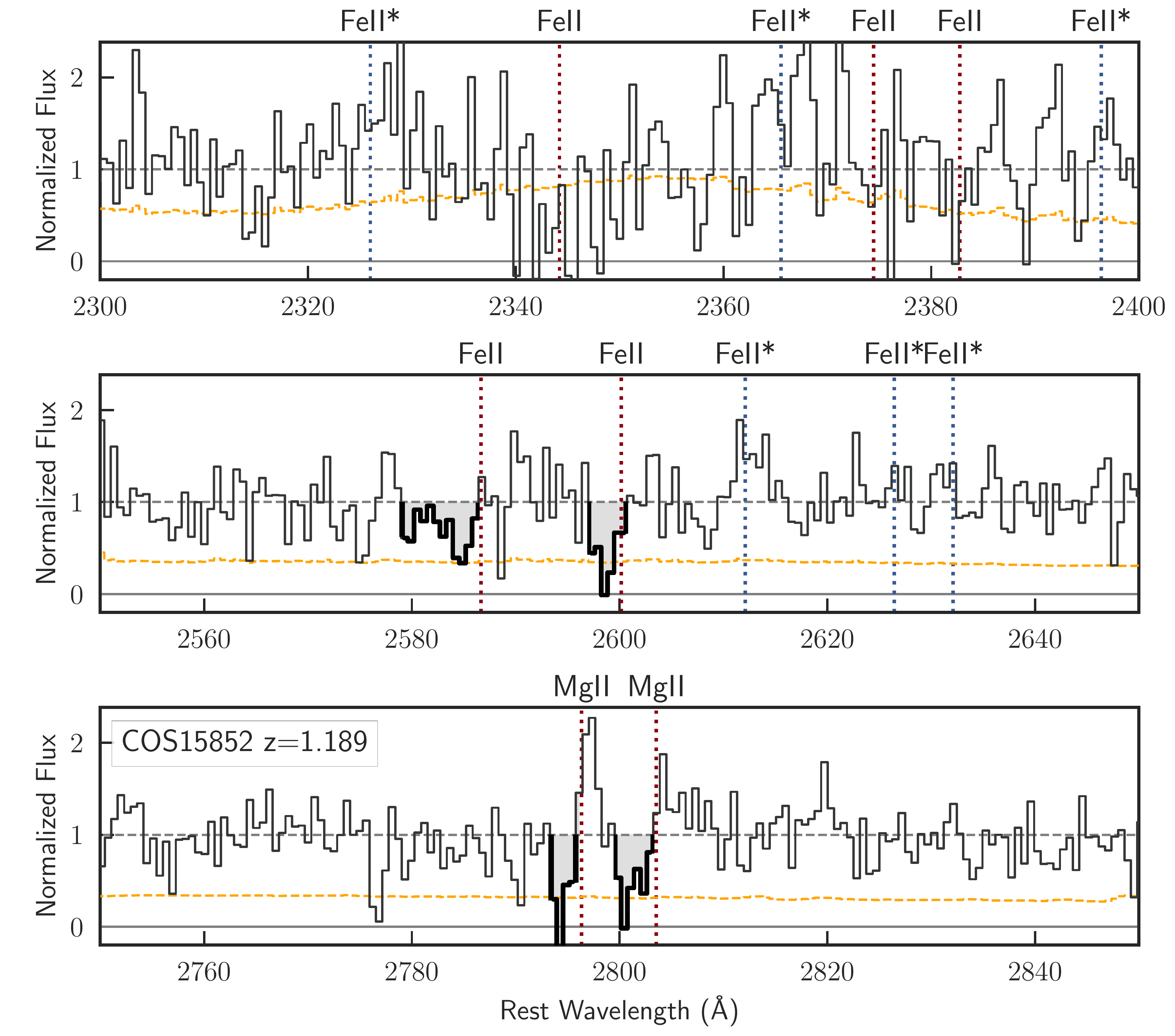}
	\end{minipage}
	\begin{minipage}{0.48\textwidth}
		\centering
		\includegraphics[width=\linewidth]{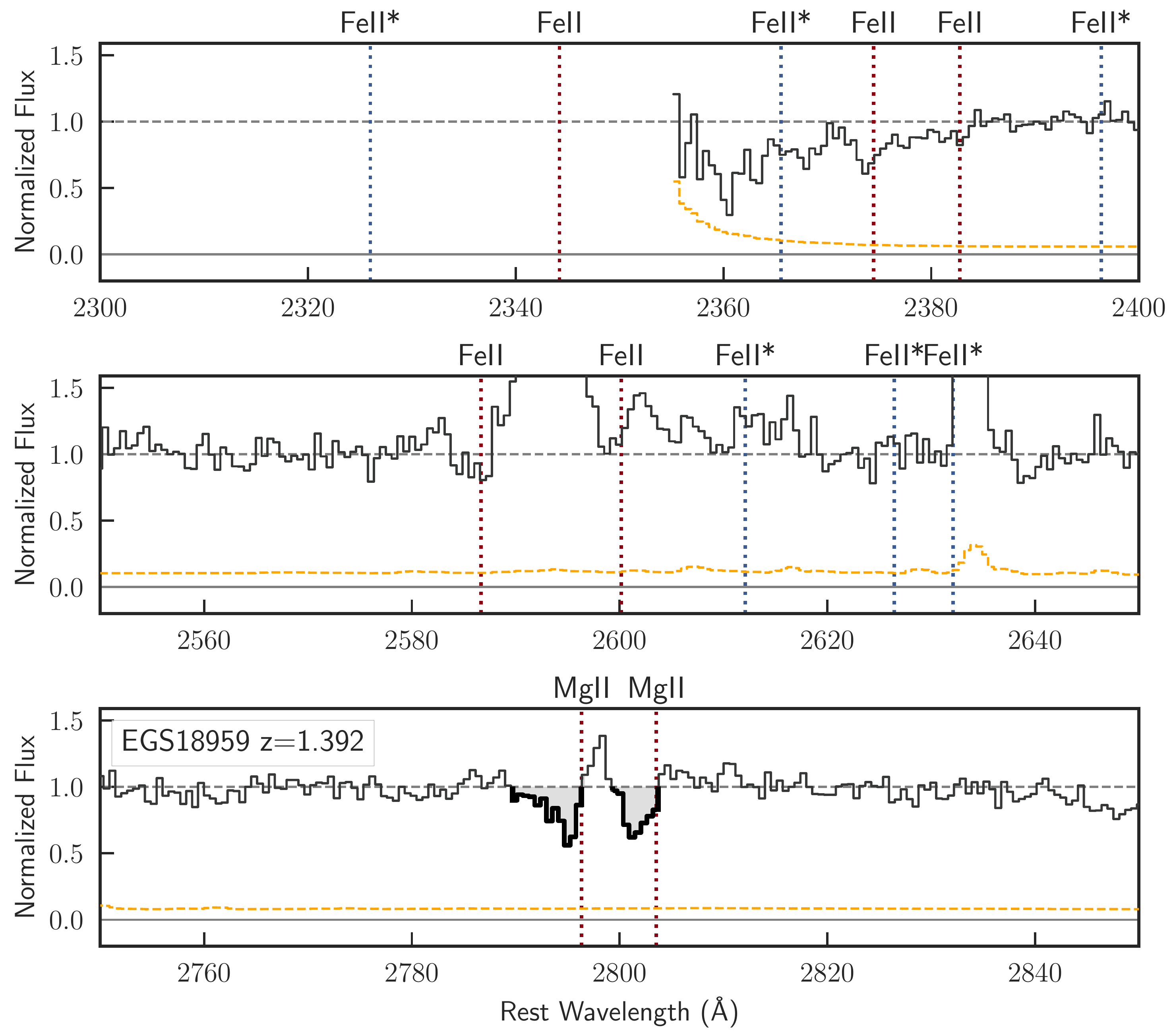}
	\end{minipage}

	\caption{Six representative DEIMOS spectra (dark gray line), with the \edit1{range over which} significantly-detected \textbf{($\geq 3\sigma$)} absorption lines \edit1{are measured} over-plotted in black \edit1{and shaded in gray}. The orange dashed line shows the total error calculated by summing the statistical and normalization uncertainties in quadrature. In the top two panels, there are only \MgII and \FeII detections respectively. In the left figure of the middle panel, there are many \FeII lines in addition to the \MgII doublet. In the middle right, there is strong \MgII emission line filling, which occurs significantly in all of the bottom four panels as well. In the bottom row, a noisy spectrum is shown with a correspondingly large error spectrum (especially in the 2300--2400 \AA\ range). On the bottom right, the galaxy has high \MgII S/N but no other lines are detected.}
	
	\label{fig:DEIMOS_spectra}
\end{figure*}

The Keck/DEIMOS spectra were reduced using the DEEP2 reduction pipeline \citep{newman_deep2_2013, cooper_spec2d_2012}\footnote{\url{https://www2.keck.hawaii.edu/inst/deimos/pipeline.html}}, which produces one-dimensional air wavelength-calibrated galaxy, inverse variance, and sky spectra. We then refined the wavelength calibration using the night sky lines, removed the instrumental signature using standard stars observed on the same night as the science data, and rebinned the spectra by a factor of two to increase the S/N; see \citet{erb_galactic_2012} for more details. The final spectra have a median S/N of 7.6 per resolution element over the wavelength range for which we measure absorption features.

We measured systemic redshifts for the sample of 22 galaxies by fitting a double Gaussian to the \OIIwave emission lines, and then shifted the spectra to the rest frame. We normalized the spectra over the wavelength range 2300--2850 \AA, as this interval contains the \FeII and \MgII spectral features of interest. \edit1{Following a slight modification of the procedure outlined by \citet{rix_spectral_2004}, we defined a series of continuum windows and fit a spline curve to the median fluxes in each of the windows. Each spectrum was then divided by the best fit to its continuum. The uncertainty on the points to be fit was defined as the standard deviation of the mean of the fluxes in each window. For a conservative estimate of the uncertainty associated with the continuum fit, we also calculated a high (and low) continuum fit by adding (subtracting) the continuum uncertainties to the median fluxes in each window. The adopted uncertainties on the continuum fits are then half of the range between the high and low continuum fits, and are typically $\approx \pm 4\%$ across the sample.} The continuum-normalized spectra were used for the analysis that follows.

We then measured the equivalent widths (EWs), velocity centroids ($\Delta v$), and maximum blueshifted velocities (\vmax) for all of the absorption lines of interest. Equivalent widths are measured by direct integration over the region for which the flux is below the continuum, and the maximum blueshifted velocity is defined as the velocity corresponding to the wavelength where the absorption feature first meets the continuum on the blue side of the line. Uncertainties in each of these quantities were determined from the 68\% confidence intervals of 1000 Monte Carlo simulations. 

Defining a detection as a $\geq3\sigma$ measurement of the equivalent width, we detect at least one \FeII line in 18 galaxies (typically \FeII $\lambda2600$, detected in 17 of 18 sources with \FeII detections). Fifteen of those 18 objects have at least two \FeII lines detected, with both \FeII $\lambda2587$ and \FeII $\lambda2600$ detected in 13 of the 15 objects with two or more \FeII detections. In four galaxies, we detect some or all of the bluer lines (\FeII $\lambda2344$, $\lambda$2374, $\lambda2383$) in addition to \FeII $\lambda2587$ and \FeII $\lambda2600$. For the \MgIIDoubletWave absorption system, \MgIIDoubletLowWave is detected in 20 objects, and 17 of those 20 have 3$\sigma$ detections of both lines. Representative examples of the spectra are shown in Figure \ref{fig:DEIMOS_spectra}.

The commonly-detected interstellar absorption lines in star-forming galaxies at $z\gtrsim1$ are typically strongly saturated, such that the equivalent width is largely determined by the covering fraction and velocity range of the absorbing gas (e.g.\ \citealt{shapley_rest-frame_2003}, \citealt{erb_galactic_2012}). In addition the interstellar medium (ISM) and outflow components of the absorption features are blended at the resolution of our spectra; we discuss this further in Section \ref{sec:composite} when looking at the line profiles of composite spectra. We confirm that for a given object the \FeII equivalent widths are generally consistent within the uncertainties, \edit1{indicating that the lines are indeed saturated. We therefore} adopt a weighted average of the equivalent widths and velocities of the detected \FeII features, using the inverse variance of each measurement as the weight and the inverse of the square root of the sum of the weights as the uncertainty in the average. For \MgII, we use only the \MgIIDoubletLowWave line for equivalent width and velocity calculations, since the blue side of \MgIIDoubletHighWave is often blended with \MgIIDoubletLowWave. Equivalent width and velocity measurements from the DEIMOS spectra are shown in Table \ref{tab:Keck}.

\section{Results}
\label{sec:results}

The complete sample of objects is shown in Figure \ref{fig:cutouts}, with F140W direct images (left) and \Ha emission line maps (right) for each galaxy. The area that maximizes the S/N of the integrated \Ha\ flux is over-plotted with a red contour. 
We find that the regions of strongest star formation are not necessarily contiguous (e.g. EGS 12858), and distinct regions with higher \Ha surface densities  may exist away from the center. From the 22 object sample, 21 objects have \SigmaSFR $>$ 0.1 \MsunPerYrPerSqrKpc, with a median SFR surface density of 0.37 \MsunPerYrPerSqrKpc.

\begin{deluxetable*}{lccccc}
	\tablecaption{Galaxy Masses and Star Formation Rates}
	
	\tablehead{\colhead{Object} & \colhead{$z_\text{sys}$\textsuperscript{a}} & \colhead{\Mstar\unskip\textsuperscript{b}} & \colhead{SFR\textsuperscript{c}} & \colhead{sSFR\textsuperscript{c}} & \colhead{\SigmaSFR\unskip\textsuperscript{c}} \\ 
		\colhead{} & \colhead{} & \colhead{($10^9$ \Msun)} & \colhead{(\MsunPerYr)} & \colhead{(G\YrInverse)} & \colhead{(\MsunPerYrPerSqrKpc)}} 
	
	\startdata
        COS 05191 & $1.22457 \pm 0.00004$ & $\phantom{1}5.2 \pm 0.8\phantom{1}$ & $15.0 \pm 1.8\phantom{1}\phantom{1}$ & $2.88 \pm 0.76$ & $0.76 \pm 0.07$ \\
        COS 09419 & $1.32996 \pm 0.00002$ & $14.3 \pm 3.1\phantom{1}$ & $30.2 \pm 5.7\phantom{1}\phantom{1}$ & $2.11 \pm 0.86$ & $0.28 \pm 0.06$ \\
        COS 10318 & $0.92615 \pm 0.00002$ & $\phantom{1}8.6 \pm 1.9\phantom{1}$ & $11.4 \pm 2.9\phantom{1}\phantom{1}$ & $1.32 \pm 0.62$ & $0.42 \pm 0.12$ \\
        COS 11696 & $1.51683 \pm 0.00004$ & $21.6 \pm 2.7\phantom{1}$ & $38.4 \pm 3.1\phantom{1}\phantom{1}$ & $1.77 \pm 0.37$ & $1.18 \pm 0.07$ \\
        COS 12589 & $1.07264 \pm 0.00001$ & $\phantom{1}4.5 \pm 1.0\phantom{1}$ & $\phantom{1}6.5 \pm 1.9\phantom{1}\phantom{1}$ & $1.45 \pm 0.75$ & $0.43 \pm 0.07$ \\
        COS 13739 & $1.21423 \pm 0.00001$ & $10.6 \pm 0.7\phantom{1}$ & $10.8 \pm 1.9\phantom{1}\phantom{1}$ & $1.02 \pm 0.24$ & $0.34 \pm 0.05$ \\
        COS 14214 & $1.40351 \pm 0.00002$ & $\phantom{1}7.2 \pm 1.4\phantom{1}$ & $19.7 \pm 3.7\phantom{1}\phantom{1}$ & $2.75 \pm 1.05$ & $0.57 \pm 0.13$ \\
        COS 15852 & $1.18903 \pm 0.00001$ & $\phantom{1}3.9 \pm 0.6\phantom{1}$ & $\phantom{1}5.1 \pm 1.2\phantom{1}\phantom{1}$ & $1.30 \pm 0.50$ & $0.23 \pm 0.02$ \\
        COS 16172 & $1.03018 \pm 0.00006$ & $73.3 \pm 21.6$ & $76.3 \pm 7.4\phantom{1}\phantom{1}$ & $1.04 \pm 0.41$ & $0.38 \pm 0.04$ \\
        COS 19180 & $1.21327 \pm 0.00001$ & $\phantom{1}2.0 \pm 0.1\phantom{1}$ & $\phantom{1}5.9 \pm 0.6\phantom{1}\phantom{1}$ & $2.97 \pm 0.45$ & $0.42 \pm 0.04$ \\
        COS 22785 & $1.23736 \pm 0.00002$ & $\phantom{1}3.8 \pm 0.7\phantom{1}$ & $\phantom{1}3.0 \pm 2.0\phantom{1}\phantom{1}$ & $0.80 \pm 0.68$ & $0.37 \pm 0.08$ \\
        COS 25161 & $1.12831 \pm 0.00001$ & $10.8 \pm 2.5\phantom{1}$ & $10.8 \pm 2.1\phantom{1}\phantom{1}$ & $1.00 \pm 0.42$ & $0.43 \pm 0.06$ \\
        COS 25927 & $1.40368 \pm 0.00004$ & $14.6 \pm 3.2\phantom{1}$ & $\phantom{1}8.0 \pm 2.2\phantom{1}\phantom{1}$ & $0.55 \pm 0.27$ & $0.30 \pm 0.03$ \\
        COS 27042 & $1.49868 \pm 0.00004$ & $24.3 \pm 1.9\phantom{1}$ & $18.1 \pm 3.2\phantom{1}\phantom{1}$ & $0.74 \pm 0.19$ & $0.67 \pm 0.06$ \\
        EGS 02986 & $1.09698 \pm 0.00002$ & $10.1 \pm 1.5\phantom{1}$ & $\phantom{1}2.1 \pm 1.0\phantom{1}\phantom{1}$ & $0.20 \pm 0.13$ & $0.13 \pm 0.02$ \\
        EGS 12858 & $1.10386 \pm 0.00001$ & $35.9 \pm 9.7\phantom{1}$ & $\phantom{1}3.2 \pm 1.1\phantom{1}\phantom{1}$ & $0.09 \pm 0.05$ & $0.07 \pm 0.01$ \\
        EGS 14380 & $1.07581 \pm 0.00005$ & $23.7 \pm 6.8\phantom{1}$ & $10.4 \pm 2.7\phantom{1}\phantom{1}$ & $0.44 \pm 0.24$ & $0.23 \pm 0.05$ \\
        EGS 18959 & $1.39209 \pm 0.00002$ & $19.3 \pm 3.4\phantom{1}$ & $\phantom{1}6.4 \pm 1.8\phantom{1}\phantom{1}$ & $0.33 \pm 0.15$ & $0.15 \pm 0.02$ \\
        EGS 24533 & $1.28145 \pm 0.00001$ & $\phantom{1}9.9 \pm 3.8\phantom{1}$ & $11.5 \pm 3.3\phantom{1}\phantom{1}$ & $1.16 \pm 0.78$ & $0.37 \pm 0.06$ \\
        EGS 26680 & $1.06009 \pm 0.00004$ & $10.6 \pm 1.6\phantom{1}$ & $\phantom{1}4.7 \pm 1.5\phantom{1}\phantom{1}$ & $0.44 \pm 0.21$ & $0.11 \pm 0.02$ \\
        EGS 27539 & $1.50368 \pm 0.00010$ & $44.2 \pm 12.5$ & $21.4 \pm 5.6\phantom{1}\phantom{1}$ & $0.48 \pm 0.26$ & $0.34 \pm 0.07$ \\
        EGS 29026 & $1.28263 \pm 0.00002$ & $15.3 \pm 5.4\phantom{1}$ & $\phantom{1}7.8 \pm 2.3\phantom{1}\phantom{1}$ & $0.51 \pm 0.33$ & $0.48 \pm 0.08$ \\
	\enddata

	\tablecomments{\\ 
		\textsuperscript{a} Systemic redshift ($z_\text{sys}$) from \OIIwave measured by Keck/DEIMOS.\\
		\textsuperscript{b} Stellar masses (\Mstar) from 3D-HST photometric catalog \citep{skelton_3d-hst_2014} \added{with uncertainties from R.\ Skelton (private communication)}, converted to a \citet{salpeter_luminosity_1955} IMF.\\
		\textsuperscript{c} Star formation rates (SFR), specific star formation rates (sSFR), and star formation rate surface densities (\SigmaSFR) computed from \textit{HST} WFC3/G141 \Ha emission line maps (see Figure \ref{fig:cutouts}).
	}
		
\end{deluxetable*}
\label{tab:HST}

\begin{deluxetable*}{lccccccc}

	\tablecaption{Absorption Line Measurements from Keck/DEIMOS}
	
	\tablehead{\colhead{Object} & \colhead{\FeII EW\textsuperscript{a}} & \colhead{\MgIIDoubletLowWave EW} & \colhead{\MgIIDoubletHighWave EW} & \colhead{\FeII $\Delta v$\textsuperscript{b}} & \colhead{\MgII $\Delta v$\textsuperscript{c}} & \colhead{\FeII \vmax\unskip\textsuperscript{d}} & \colhead{\MgII \vmax\unskip\textsuperscript{e}} \\ 
		\colhead{} & \colhead{(\AA)} & \colhead{(\AA)} & \colhead{(\AA)} & \colhead{(\kms)} & \colhead{(\kms)} & \colhead{(\kms)} & \colhead{(\kms)}} 
	
	\startdata
		COS 05191 & $3.8 \pm 0.6$ & $3.1 \pm 0.5$ & $2.0 \pm 0.5$ & \phantom{1}$-96 \pm 52$ & $-333 \pm 30$ & $-527 \pm 56$\phantom{0} & $-612 \pm 33$\phantom{0} \\
		COS 09419 & $2.5 \pm 0.5$ & $1.6 \pm 0.4$ & $3.5 \pm 0.5$ & $-118 \pm 49$ & $-152 \pm 34$ & $-432 \pm 119$ & $-475 \pm 94$\phantom{0} \\
		COS 10318 & \nodata & $3.3 \pm 0.5$ & $3.7 \pm 0.5$ & \nodata & \phantom{1}$-53 \pm 51$ & \nodata & $-766 \pm 189$ \\
		COS 11696 & $2.9 \pm 0.5$ & $4.0 \pm 0.6$ & $3.0 \pm 0.4$ & $-136 \pm 54$ & $-325 \pm 43$ & $-454 \pm 79$\phantom{0} & $-825 \pm 116$ \\
		COS 12589 & $2.4 \pm 0.4$ & \nodata & \nodata & \phantom{1}\phantom{0}$-8 \pm 32$ & \nodata & $-378 \pm 54$\phantom{0} & \nodata \\
		COS 13739 & $2.2 \pm 0.4$ & $2.4 \pm 0.6$ & $3.0 \pm 0.7$ & \phantom{1}\phantom{0}$\phantom{-}7 \pm 43$ & \phantom{1}$-71 \pm 82$ & $-268 \pm 56$\phantom{0} & $-520 \pm 230$ \\
		COS 14214 & $3.2 \pm 0.2$ & $4.3 \pm 0.4$ & $3.6 \pm 0.5$ & \phantom{1}\phantom{0}$\phantom{-}1 \pm 17$ & \phantom{1}$-73 \pm 21$ & $-392 \pm 49$\phantom{0} & $-614 \pm 61$\phantom{0} \\
		COS 15852 & $2.1 \pm 0.5$ & $1.9 \pm 0.5$ & $2.0 \pm 0.6$ & $-222 \pm 46$ & $-202 \pm 28$ & $-494 \pm 144$ & $-359 \pm 67$\phantom{0} \\
		COS 16172 & $2.8 \pm 0.4$ & $3.3 \pm 0.5$ & $3.6 \pm 0.5$ & \phantom{1}$\phantom{-}69 \pm 38$ & \phantom{1}$-44 \pm 34$ & $-300 \pm 70$\phantom{0} & $-416 \pm 72$\phantom{0} \\
		COS 19180 & $2.2 \pm 0.5$ & \nodata & \nodata & \phantom{1}\phantom{0}$-9 \pm 46$ & \nodata & $-294 \pm 81$\phantom{0} & \nodata \\
		COS 22785 & $1.6 \pm 0.5$ & $1.5 \pm 0.4$ & \nodata & \phantom{1}$-66 \pm 53$ & $-228 \pm 46$ & $-423 \pm 70$\phantom{0} & $-520 \pm 130$ \\
		COS 25161 & $2.8 \pm 0.7$ & $3.1 \pm 0.6$ & \nodata & $-278 \pm 33$ & $-253 \pm 30$ & $-638 \pm 42$\phantom{0} & $-557 \pm 0$\phantom{0}\phantom{0} \\
		COS 25927 & $2.3 \pm 0.4$ & $1.8 \pm 0.4$ & $1.9 \pm 0.6$ & \phantom{1}$-41 \pm 30$ & \phantom{1}$-52 \pm 38$ & $-288 \pm 58$\phantom{0} & $-303 \pm 91$\phantom{0} \\
		COS 27042 & $4.0 \pm 0.5$ & $3.4 \pm 1.0$ & $2.9 \pm 0.8$ & $-149 \pm 27$ & $-223 \pm 66$ & $-518 \pm 43$\phantom{0} & $-654 \pm 175$ \\
		EGS 02986 & $2.1 \pm 0.7$ & $1.6 \pm 0.4$ & $1.8 \pm 0.4$ & \phantom{1}$\phantom{-}87 \pm 69$ & $\phantom{-}107 \pm 26$ & $-295 \pm 150$ & $-100 \pm 35$\phantom{0} \\
		EGS 12858 & $2.0 \pm 0.4$ & $2.7 \pm 0.6$ & \nodata & \phantom{1}$\phantom{-}78 \pm 38$ & $-207 \pm 42$ & $-306 \pm 42$\phantom{0} & $-534 \pm 69$\phantom{0} \\
		EGS 14380 & \nodata & $1.9 \pm 0.4$ & $2.0 \pm 0.4$ & \nodata & \phantom{1}$-35 \pm 26$ & \nodata & $-297 \pm 70$\phantom{0} \\
		EGS 18959 & \nodata & $1.2 \pm 0.2$ & $1.0 \pm 0.1$ & \nodata & $-262 \pm 46$ & \nodata & $-760 \pm 183$ \\
		EGS 24533 & $1.8 \pm 0.1$ & $2.9 \pm 0.3$ & $2.3 \pm 0.2$ & \phantom{1}$-79 \pm 18$ & $-204 \pm 40$ & $-429 \pm 51$\phantom{0} & $-756 \pm 128$ \\
		EGS 26680 & \nodata & $2.9 \pm 0.4$ & $2.4 \pm 0.4$ & \nodata & \phantom{1}$-82 \pm 24$ & \nodata & $-442 \pm 35$\phantom{0} \\
		EGS 27539 & $2.9 \pm 0.4$ & $3.8 \pm 0.4$ & $3.9 \pm 0.4$ & \phantom{1}$-51 \pm 29$ & $-214 \pm 22$ & $-411 \pm 35$\phantom{0} & $-565 \pm 29$\phantom{0} \\
		EGS 29026 & $3.2 \pm 0.5$ & $3.2 \pm 0.5$ & $4.5 \pm 0.5$ & \phantom{1}$\phantom{-}15 \pm 21$ & \phantom{1}$-48 \pm 42$ & $-249 \pm 34$\phantom{0} & $-632 \pm 192$ \\
	\enddata

	\tablecomments{\\ 
		\textsuperscript{a} Weighted average of equivalent widths from all significantly detected \FeII absorption lines.\\
		\textsuperscript{b} Weighted average of \FeII centroid velocities from all significantly detected \FeII lines.\\
		\textsuperscript{c} \MgII centroid velocities computed using \MgIIDoubletLowWave.\\
		\textsuperscript{d} Weighted average of \FeII maximum velocities from all significantly detected \FeII lines.\\
		\textsuperscript{e} \MgII maximum velocities computed using \MgIIDoubletLowWave.\\
	}

\end{deluxetable*}
\label{tab:Keck}

We test for the presence of outflows or inflows by determining whether the centroids of the \FeII absorption lines are consistent with the systemic velocity. At the 1$\sigma$ level, 14 objects have outflows and three have inflows, while the remaining five objects have velocity offsets consistent with zero. Requiring a 3$\sigma$ threshold results in six objects with outflows and no inflows. K-S tests indicate that each of these subsamples is consistent with being drawn from the same parent distribution in terms of stellar mass, SFR, and SFR surface density. We use the \FeII centroid velocities because the \FeII absorption lines are less susceptible to emission line filling. For the four galaxies with \FeII non-detections, we use the \MgIIDoubletLowWave centroid instead. Four of the six objects with 3$\sigma$ detections of outflows are measured with \FeII, indicating that outflow detections are not dominated by the effects of \MgII emission line filling.

\begin{figure*}
	\includegraphics[width=\textwidth]{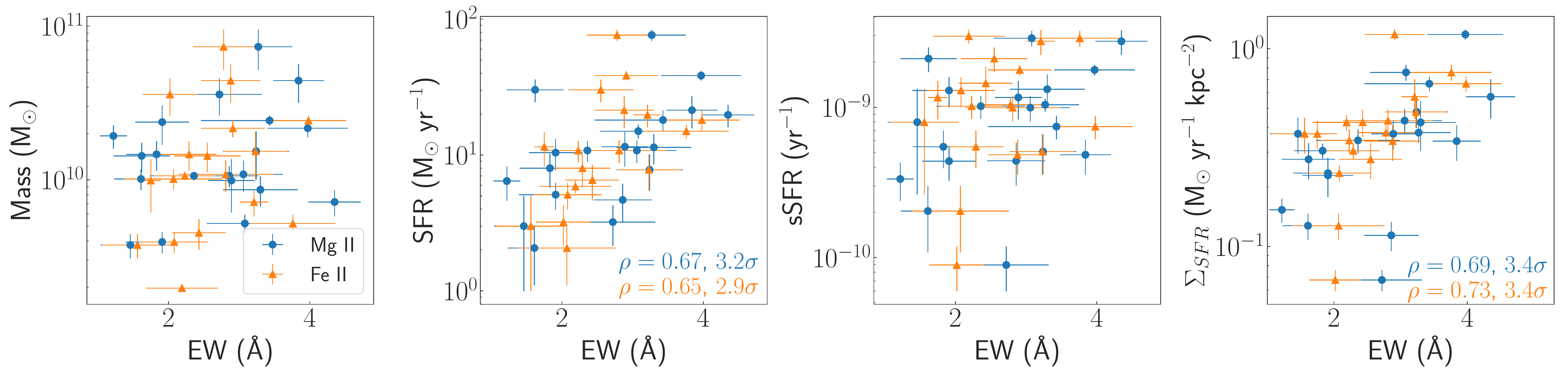}
	\caption{From left to right, we plot stellar mass, star formation rate, specific star formation rate, and star formation rate surface density against \FeII (orange triangles) and \MgIIDoubletLowWave (blue circles) equivalent widths. As described in the text, there are significant correlations between \MgII EW and SFR, \MgII EW and \SigmaSFR, and \FeII EW and \SigmaSFR, and a marginal correlation between \FeII EW and SFR as well. \added{Spearman correlation coefficients and significances are shown in the plots for which there are significant correlations, color coded by absorption line.}}
	\label{fig:ew}
\end{figure*}

\begin{figure*}
	\includegraphics[width=\textwidth]{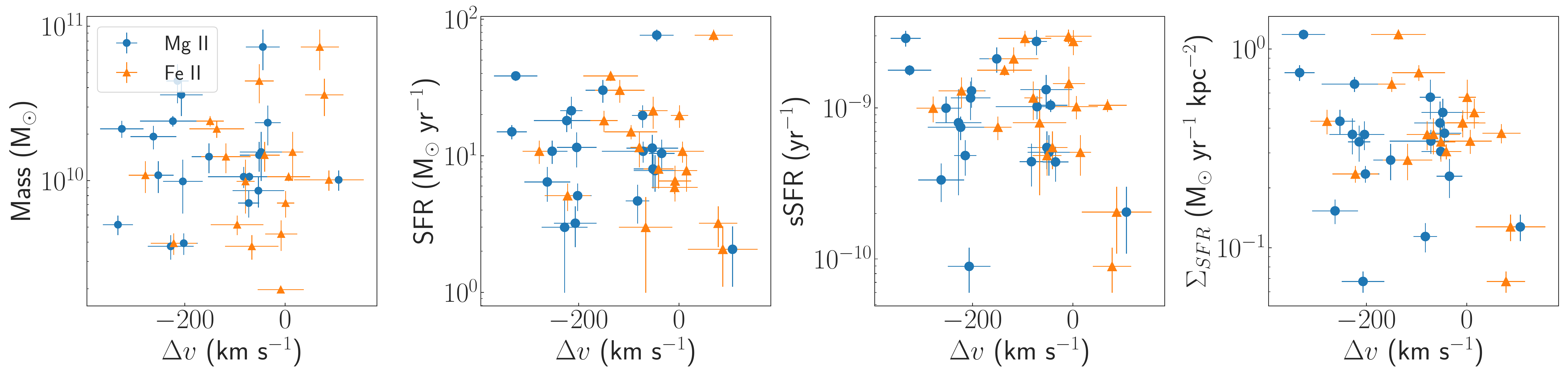}
	\caption{From left to right, we plot stellar mass, star formation rate, specific star formation rate, and star formation rate surface density vs.\ the \FeII and \MgII velocity centroids $\Delta v$. Symbols are as in Figure \ref{fig:ew}, and we find no significant correlations between any of the quantities ($\sigma\leq 1.4$).}
	\label{fig:deltav}
\end{figure*}

\begin{figure*}
	\includegraphics[width=\textwidth]{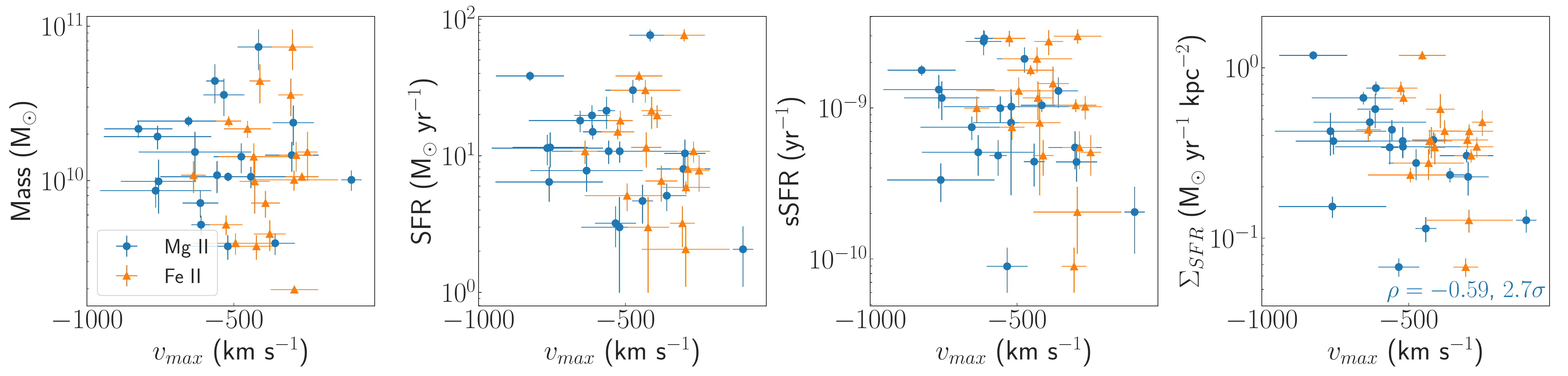}
	\caption{From left to right, we plot stellar mass, star formation rate, specific star formation rate, and star formation rate surface density vs.\ the \FeII and \MgII maximum velocity \vmax, defined as the velocity where the absorption line reaches the continuum on the blue side of the line. The symbols remain the same as in Figures \ref{fig:ew} and \ref{fig:deltav}. There is a marginal correlation between SFR surface density and \MgII maximum outflow velocity \added{(correlation coefficent and significance shown in the figure)}, and no other correlations are found ($\sigma\leq 1.9$).}
	\label{fig:vmax}
\end{figure*}

In Figures \ref{fig:ew}, \ref{fig:deltav}, and \ref{fig:vmax} we compare the stellar mass, SFR, sSFR, and \SigmaSFR against EW, $\Delta v$, and \vmax respectively for all objects in the dataset. We test for correlations using the Spearman correlation coefficient $\rho$, and determine the significance, $\sigma$, which represents the number of standard deviations from the null hypothesis (no correlation between quantities). We find that the strength of \FeII and \MgII absorption increases
with the star formation rate surface density:\ both the \FeII and \MgII equivalent widths are correlated with \SigmaSFR at the 3.4$\sigma$ level, with similar correlation coefficients of $\rho=0.73$ and $\rho=0.69$ respectively.

We also that find \MgII equivalent width is positively correlated ($\rho=0.67$) with the star formation rate at the 3.2$\sigma$ level. In addition, \FeII EW has a marginal correlation ($\rho=0.65$, $\sigma=2.9$) with the SFR. These correlations suggest that galaxies with more star formation drive more gas to a broader range of velocities, thereby increasing the equivalent width of the absorption lines. Finally, we find a weak correlation between \MgII maximum outflow velocity and the SFR surface density ($\rho=-0.59$), with 2.7$\sigma$ significance. We find no correlations between between outflow velocity and the sSFR, nor any correlations between stellar mass and EW or outflow velocity. 

\added{For a more direct comparison with the literature, we test the SFR surface densities $\Sigma_{\rm{SFR,\, B}}$ computed using the \citet{bordoloi_dependence_2014} method (discussed in Section \ref{sec:sfr_sd}) for correlations with our outflow-related quantities. The significant correlations between \FeII and \MgII EW and the SFR surface density discussed above are not present when $\Sigma_{\rm{SFR,\, B}}$ is used; we find EW and $\Delta v$ to be uncorrelated ($\sigma\leq 1.9$) with $\Sigma_{\rm{SFR,\, B}}$ for both \FeII and \MgII. On the other hand, we observe a $3.5\sigma$ correlation between \MgII maximum outflow velocity and $\Sigma_{\rm{SFR,\, B}}$, slightly more significant than
the $2.7\sigma$ correlation seen above using our optimal aperture method.}

The lack of correlation between equivalent width and stellar mass provides further insight into the correlations described above. If the EW is dominated by the ISM component, we would expect to observe an increase in equivalent width with increasing mass and velocity dispersion; since we do not see this increase, we conclude that changes in the \MgII and \FeII EW across the sample are likely to be due more to the outflow component than the ISM. The \MgII and \FeII EW vs.\ SFR correlations we detect are also then likely to be primarily determined by the outflow.  
We discuss the relative contributions of the ISM and outflow components further in Section \ref{sec:composite} below.

We also tested subsets of the 22 object sample for correlations between the stellar population parameters and the outflow quantities. In particular, we calculated Spearman correlation coefficients and determined their significance for the 14 object and 6 object subsamples with 1$\sigma$ and 3$\sigma$ outflows respectively. In both cases, no significant correlations ($\sigma \leq 2.6$) were detected between any of the quantities, probably due to the small number of objects in these subsets.

Below, we discuss the results from composite absorption line spectra constructed from low and high subsamples of galaxy physical properties.

\subsection{Composite Spectra}
\label{sec:composite}

\begin{figure*}
	\plotone{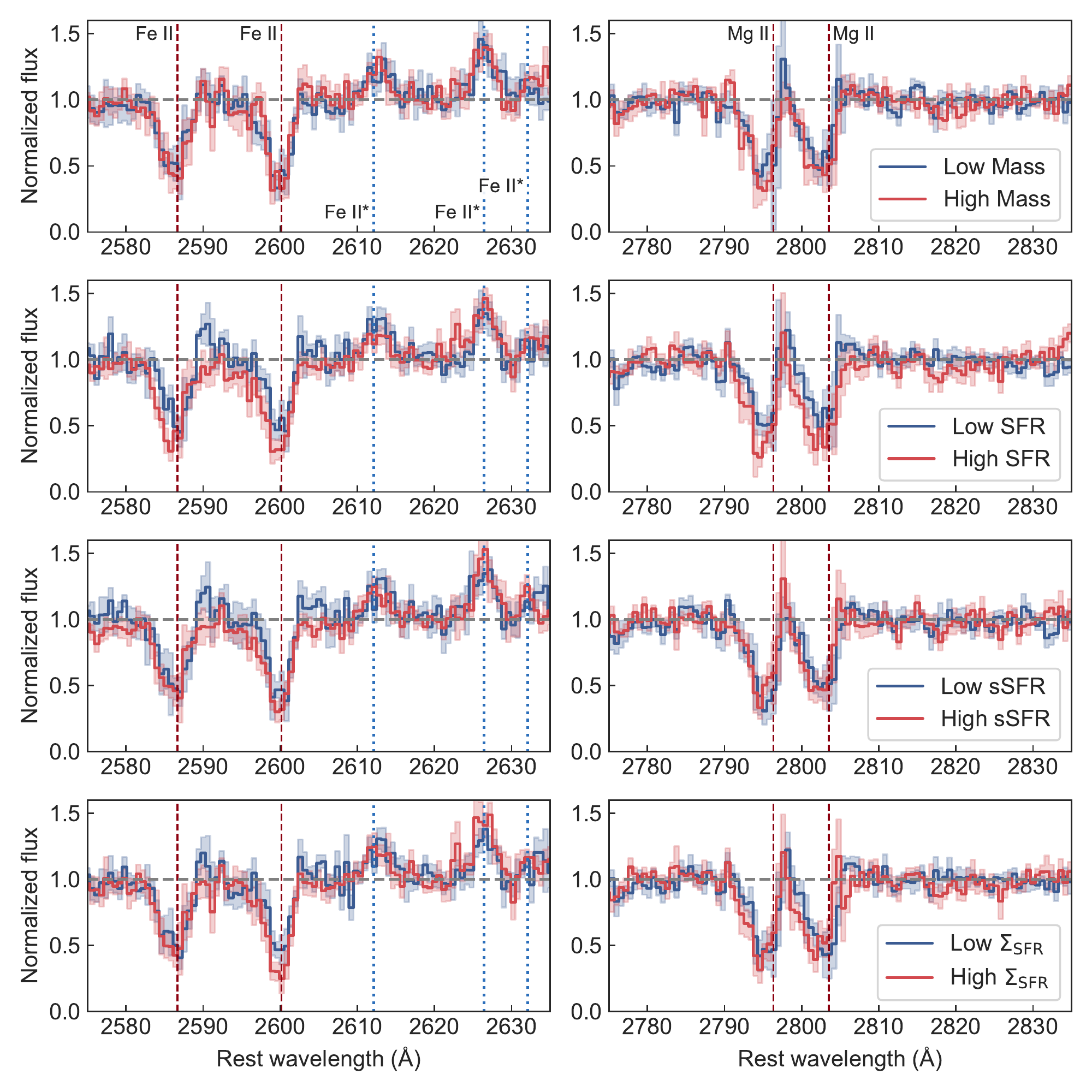}

	\caption{Composite spectra constructed from subsets of galaxies based on the galaxy properties of Figure \ref{fig:hist}. The left column of panels shows \FeII transitions at $\sim 2600$ \AA, while the right column shows the \MgIIDoubletWave doublet. From top to bottom, composite spectra are binned by stellar mass, SFR, specific SFR, and SFR surface density. The color scheme is the same as Figure \ref{fig:hist}, with red and blue corresponding to the low and high respective subsamples of a particular quantity. Uncertainties are shown as faded bars behind the spectra.}
	\label{fig:composite}
\end{figure*}

\begin{deluxetable*}{lcccccc}

	\tablecaption{Composite Spectra Properties}

	\tablehead{\colhead{Composite\textsuperscript{a}} & \colhead{\FeII EW\textsuperscript{b}} & \colhead{\MgIIDoubletLowWave EW} & \colhead{\FeII $\Delta v$\textsuperscript{c}} & \colhead{\MgII $\Delta v$\textsuperscript{d}} & \colhead{\FeII \vmax\unskip\textsuperscript{e}} & \colhead{\MgII \vmax\unskip\textsuperscript{f}} \\ 
		\colhead{} & \colhead{(\AA)} & \colhead{(\AA)} & \colhead{(\kms)} & \colhead{(\kms)} & \colhead{(\kms)} & \colhead{(\kms)}} 
	
	\startdata
	    Low Mass & $2.2\pm0.2$ & $2.1\pm0.5$ & \phantom{1}$-76\pm22$ & $-198\pm46$ & $-498\pm49$ & $-574\pm96$ \\
        High Mass & $2.2\pm0.2$ & $2.3\pm0.3$ & \phantom{1}$-48\pm20$ & $-179\pm28$ & $-348\pm34$ & $-574\pm32$ \\
        Low SFR & $1.7\pm0.2$ & $1.7\pm0.3$ & \phantom{1}$-25\pm18$ & $-173\pm31$ & $-384\pm31$ & $-509\pm32$ \\
        High SFR & $3.0\pm0.3$ & $2.6\pm0.3$ & $-102\pm35$ & $-191\pm22$ & $-519\pm66$ & $-574\pm64$ \\
        Low sSFR & $2.0\pm0.2$ & $2.3\pm0.3$ & \phantom{1}$-44\pm22$ & $-156\pm32$ & $-394\pm49$ & $-509\pm64$ \\
        High sSFR & $2.7\pm0.2$ & $2.2\pm0.3$ & $-104\pm30$ & $-200\pm33$ & $-470\pm64$ & $-574\pm96$ \\
        Low \SigmaSFR & $1.9\pm0.2$ & $2.0\pm0.3$ & \phantom{1}$-44\pm22$ & $-145\pm28$ & $-422\pm58$ & $-509\pm32$ \\
        High \SigmaSFR & $2.6\pm0.2$ & $2.3\pm0.3$ & \phantom{1}$-86\pm26$ & $-193\pm25$ & $-467\pm62$ & $-574\pm64$ \\
	\enddata

	\tablecomments{\\ 
	    \textsuperscript{a} See Figure \ref{fig:hist} for distributions and ranges of each subsample.\\
		\textsuperscript{b} Weighted average of \FeII $\lambda2587$ and \FeII $\lambda2600$ equivalent widths.\\
		\textsuperscript{c} Weighted average of \FeII centroid velocities from \FeII $\lambda2587$ and \FeII $\lambda2600$.\\
		\textsuperscript{d} \MgII centroid velocities computed using \MgIIDoubletLowWave.\\
		\textsuperscript{e} Weighted average of \FeII maximum velocities from \FeII $\lambda2587$ and \FeII $\lambda2600$.\\
		\textsuperscript{f} \MgII maximum velocities computed using \MgIIDoubletLowWave.\\
	}

\end{deluxetable*}
\label{tab:composite}

We created composite spectra by dividing the sample in half by stellar mass, SFR, sSFR, and SFR surface density and constructing the median spectrum of each subsample  (we refer to the subsamples as low mass, high mass, low SFR, etc.). The spectra are shown Figure \ref{fig:composite}, with error spectra computed from the standard deviations of 100 bootstrap resamples of each subsample. The distributions of the low and high subsets are shown by the blue and red histograms respectively in Figure \ref{fig:hist}. For the low (high) subsamples, we find a median stellar mass of $7.2\times 10^{9}$ \Msun ($2.2\times 10^{10}$ \Msun), a median SFR of 5.9 \MsunPerYr (18.1 \MsunPerYr), a median specific SFR of 0.5 G\YrInverse (1.4 G\YrInverse), and a median SFR surface density of 0.23 \MsunPerYrPerSqrKpc (0.43 \MsunPerYrPerSqrKpc). The difference between the medians of the low and high stellar mass, SFR, and sSFR composites is a factor of $\sim 3$, while the medians of the low and high \SigmaSFR samples differ by a factor of $\sim 2$. \added{We also note that the separation between the bins is larger than the uncertainties on the binned quantities by factors ranging from 2.3 (sSFR) to 6.7 (stellar mass).} 

For each physical quantity, we measured equivalent widths, centroid velocities, and maximum velocities for both of the composite spectra, using only the \FeII $\lambda2587$, \FeII $\lambda2600$, and \MgIIDoubletWave lines because the lines at redder wavelengths are covered by most (21/22) or all of the individual spectra. We take the weighted average of the \FeII $\lambda2587$ and \FeII $\lambda2600$ lines for the most direct comparison to the individual spectra. The absorption line measurements from the composite spectra are reported in Table \ref{tab:composite}.

To see how the outflow parameters change with galaxy physical properties, we compare the difference in equivalent widths, centroid velocities, and maximum velocities for each low and high subsample (e.g.\ the \FeII high mass EW vs. \FeII low mass EW). Overall, we see the most significant differences between subsets when measuring the \FeII absorption features, \added{likely because the use of the weighted average of the \FeII $\lambda2587$ and \FeII $\lambda2600$ lines decreases the uncertainties}. In contrast, the \MgII EW and velocity differences between low and high subsamples tend to remain small with respect to the \FeII measurements. This difference may be because we use only \MgIIDoubletLowWave for the calculations (as with the individual spectra) which \added{tends to have higher uncertainties with respect to the \FeII weighted averages and} is complicated by emission line filling.

On that note, the most significant difference between the composites is found when comparing the \FeII equivalent widths of the high and low star-formation-related quantities (SFR, sSFR, and \SigmaSFR), with the two SFR subsamples showing the largest difference of $1.3\pm0.3$ \AA. We find almost no difference between the low and high stellar mass subsets across all three absorption line measurements, which suggests that the changes in the velocity and equivalent width of the specific star formation rate subsamples are primarily due to changes in the SFR rather than the mass. Excluding the low and high SFR subsamples, the \MgII EWs also have differences consistent with zero. These results roughly support the \FeII EW vs.\ SFR correlation as well as the \FeII EW vs.\ \SigmaSFR correlation we found by looking at the individual objects.

In comparing the outflow velocities between the low and high subsamples, we again
find that the \FeII measurements vary more between subsamples than the \MgII velocities. The \FeII centroid velocities are consistently faster in all of the high SFR-related composites, with the largest difference of $-77\pm39$ \kms\ found between the low and high SFR subsamples. We also find a significantly faster \FeII maximum velocity in the high SFR sample. The only significant velocity-related difference in the \MgII composite measurements is found for the SFR surface densities, for which the centroid velocity is faster in the higher \SigmaSFR sample.

By looking at the profiles of the composite spectra, we can also try to see how the ISM and outflow components of the \MgII and \FeII absorption lines change between subsamples. In the second row of Figure \ref{fig:composite} (the composites based on SFR, which show the largest differences between high and low subsamples), we note that the red wings of the absorption lines have very similar profiles. This suggests that the velocity dispersion of the ISM is similar in the two subsets, although the spectral resolution of $\sim 180$ \kms\ likely prevents us from fully resolving the ISM component. Larger differences are found on the blue sides of the lines, implying that the difference between the  spectra of the two subsamples is largely determined by the outflow component. This, along with the higher covering fractions indicated by the larger depth of the lines throughout the absorption profile in the high SFR subsample (especially on the blue sides of the lines), points to stronger outflow components associated with higher SFRs.

We summarize our study and discuss the results and future prospects below.

\section{Summary and Discussion}
\label{sec:summary}

We have investigated the relationship between galactic outflows and star formation by compiling a joint dataset of 22 star-forming galaxies at $1\lesssim z\lesssim 1.5$ with rest-frame near-UV absorption line spectra from Keck/DEIMOS and \Ha emission line maps from WFC3/G141 grism spectra taken as part of the 3D-HST survey. All of the objects have at least a 3$\sigma$ detection of the \Ha emission line; 18 have \FeII absorption, and 20 have \MgII significantly detected. \edit1{The sample has a median mass and standard deviation of $\log($\Mstar/\Msun\unskip$)=10.2 \pm 0.5$}, and a median redshift of $1.22$ with standard deviation $0.16$. Our primary results are enumerated below.

\begin{enumerate}
	\item
	We used the grism data and the \textsc{Grizli} pipeline to construct 1D and 2D spectra (see Figure \ref{fig:2d-spec} for sample 2D spectra) and \Ha emission line maps for each of the 22 objects. We determined the sizes of the regions of strongest star formation from the \Ha maps, choosing the area that maximizes the S/N of the integrated \Ha flux (Figure \ref{fig:cutouts}). We explained this method of area measurement in detail in Section \ref{sec:sfr_sd}. With the \Ha luminosities and sizes, we compute star formation rates, specific star formation rates, and star formation rate surface densities. These quantities and their respective distributions are listed in Table \ref{tab:HST} and plotted in Figure \ref{fig:hist}.
	
	\item From the DEIMOS spectra, we measured \FeII and \MgII equivalent widths, centroid velocities, and maximum velocities (the velocity where the flux reaches the continuum on the blue side of the line) for each of the objects. These measurements are given in Table \ref{tab:Keck}, and a representative sample of the spectra is shown in Figure \ref{fig:DEIMOS_spectra}.
	
	\item The results from the \Ha and absorption line measurements	are combined in Figures \ref{fig:ew}, \ref{fig:deltav}, and \ref{fig:vmax}, in which we plot stellar mass, SFR, sSFR, and \SigmaSFR against EW, centroid velocity ($\Delta v$), and maximum velocity (\vmax) respectively. Spearman correlation coefficients and their significances are computed for each of the relationships. The \FeII and \MgII equivalent widths are both positively correlated with the star formation rate surface density at the $\sim 3.4\sigma$ level, and the \MgII equivalent width also increases with the star formation rate with $\sigma=3.2$. Marginal correlations are found between \FeII EW and the SFR ($\sigma=2.9$), and \MgII maximum outflow velocity versus SFR surface density ($\sigma=2.7$). There are no significant correlations ($\sigma<2$) between the other plotted quantities.
	
	\item Composite spectra were formed by splitting the dataset into low and high subsamples based on stellar mass, SFR, sSFR, and SFR surface density (Figure \ref{fig:composite}). For each low and high subset, we measured \FeII and \MgII equivalent widths, centroid velocities, and maximum velocities (see Table \ref{tab:composite}). For all of the star-formation-related quantities (SFR, sSFR, and \SigmaSFR), the \FeII absorption lines show significantly larger equivalent widths and centroid velocities for the high subsets relative to the low subsamples. We find that the \FeII and \MgII absorption lines in the high SFR composites have stronger blue wings, supporting the hypothesis that the increase in EW seen with SFR in the spectra of individual objects is due to an increase in the strength of the outflow component with SFR.
	
\end{enumerate}

Given the complex physics of galactic outflows, many authors have attempted to constrain their driving mechanisms by identifying relationships between galactic properties (e.g.\ mass, SFR) and outflow-related quantities such as wind velocity. In the local universe ($z\sim 0$), \citet{martin_mapping_2005}, \citet{rupke_outflows_2005}, \added{\citet{chisholm_scaling_2015}}, and \citet{heckman_systematic_2015} have observed outflows in relatively small ($\sim 50$) samples of star-forming galaxies. These samples cover a wide range in mass ($\sim10^7$--$10^{11}$ \Msun) and SFR ($\sim0.1$ to nearly 1000 \MsunPerYr), and the inclusion of dwarf galaxies with low masses and star formation rates has been crucial to the detection of trends between galaxy and outflow properties. Multiple studies have extended these observations to $z\sim1$ (e.g.\  \citealt{weiner_ubiquitous_2009}, \citealt{rubin_persistence_2010}, \citealt{kornei_properties_2012}, \citealt{martin_demographics_2012}, \citealt{bordoloi_dependence_2014}, \added{\citealt{rubin_evidence_2014}}); however, the higher redshift samples span only the upper end of parameter space (stellar masses $\gtrsim10^9$ \Msun, SFRs $\gtrsim1$–$10$ \MsunPerYr) and often rely on the coadding of large numbers of spectra.

From these observations, several trends have been detected that link star formation to outflow characteristics. Various studies have detected a shallow increase in outflow velocity with star formation rate, finding relationships similar to $v\sim\text{SFR}^{0.3}$ \edit1{\citep{martin_mapping_2005,rupke_outflows_2005,weiner_ubiquitous_2009, chisholm_scaling_2015, trainor_spectroscopic_2015}}. Outflow velocities have also been found to \edit1{generally} increase with SFR surface density, both locally \citep{heckman_systematic_2015} and at higher redshifts \citep{kornei_properties_2012}, \edit1{although this is not always observed \citep{chisholm_scaling_2015,rubin_evidence_2014}}. Most studies of galactic outflows at $z\sim1$ have observed correlations between \FeII and \MgII equivalent widths and stellar mass, SFR, or SFR surface density \edit1{\citep{martin_demographics_2012,weiner_ubiquitous_2009,rubin_persistence_2010,bordoloi_dependence_2014,rubin_evidence_2014}}. Given that the observed \MgII and \FeII transitions are optically thick, the equivalent width is primarily determined by the covering fraction and velocity distribution of the interstellar gas (with a potential additional contribution from emission filling in the case of \MgII). The observed correlations with equivalent width then support a general model in which galaxies with more intense star formation drive outflows with a higher covering fraction to a wider range of velocities.

Our results are consistent with this model and with previous studies at $z\sim1$ in the sense that the strongest correlations we observe between individual objects are related to the \FeII and \MgII equivalent width, which increase with SFR and SFR surface density. The composite spectra in particular highlight the relationship between star formation and outflows, with larger \FeII equivalent widths and higher velocities seen in the higher subset of all the spectra based on SFR-related quantities. These findings echo previous results at $z\sim1$ which observe ties between the SFR and SFR surface density and outflow velocity \citep{kornei_properties_2012,heckman_systematic_2015,heckman_implications_2016}, confirming the close connection between feedback and star formation.

The sample studied here spans slightly more than an order of magnitude in stellar mass, from $2$ -- $70\times10^9$ \Msun, and within this relatively narrow range we find no correlations with outflow properties measured from individual spectra. Most of the composite spectra also show no differences in velocity when divided by mass, although we do find that the \FeII maximum outflow velocity is higher in the low mass subsample, perhaps contrary to expectations. The lack of a relationship between mass and outflow velocity in the individual objects may be unsurprising given previous findings that these quantities are only weakly correlated \edit1{\citep{heckman_systematic_2015,chisholm_scaling_2015}}. The lack of correlation between mass and equivalent width in both the individual and composite spectra may be more unexpected, since most other studies at $z\sim1$ have observed that EW increases with stellar mass \edit1{\citep{weiner_ubiquitous_2009,rubin_persistence_2010,martin_demographics_2012,kornei_properties_2012,bordoloi_dependence_2014,rubin_evidence_2014}}; however, these studies are based on larger samples covering a wider range in mass.

We can make a more quantitative comparison between our sample and previous results by assuming the scaling relation \vmax$\sim\text{SFR}^{0.3}$, as found in previous studies. Using the median values of \vmax and SFR in our sample to normalize the relationship, our full range of SFRs then predicts a velocity range of \vmax~$\sim250$--$700$ \kms\ across the full sample, much wider than observed; however, two-thirds of our sample lies in the narrower range of $5<{\rm{SFR}}$/(\MsunPerYr) $<22$, and within this range the observed scatter in \vmax is comparable to the expected variation due to SFR. This suggests that the intrinsic scatter in the velocity--SFR relationship is too large for it to be detected within the range of star formation rates probed here. We note that although \citet{heckman_systematic_2015} observe a strong correlation between outflow velocity and SFR in the local universe, there is nearly an order of magnitude variation in outflow velocity at a given SFR at the upper end of their sample.

Although our small range of star formation rates likely prevents us from finding a correlation between outflow velocity and SFR in the individual objects, we do find
a relationship between \FeII maximum outflow velocity and SFR in the composite spectra that is consistent with previous studies:\ using the median SFRs for the low and high SFR subsets along with their respective maximum velocities, we find that \FeII \vmax $\sim$~SFR$^{0.27}$. Since composite spectra improve the measurement of weak features, the maximum outflow velocity can be measured more robustly, which may explain why this correlation is not seen in the individual objects. We also note that using the \FeII centroid velocities $\Delta v$ rather than the maximum velocities \vmax results in the significantly stronger relationship  $\Delta v \sim$~SFR$^{1.25}$.

Some of the objects in our sample do not follow the general trends between outflow and galaxy properties we have discussed. In particular, we focus on the galaxies which have high star formation rate surface densities but low outflow velocities. Our dataset has four objects (COS 12589, COS 14214, COS 19180, and EGS 29026) which have centroid velocities consistent with zero and star formation rate surface densities above the sample median of 0.4 \MsunPerYrPerSqrKpc. One possible explanation for this may be that outflows are present, but are collimated and pointed away from our line of sight; evidence for non-spherical outflow geometries has been found in other studies at similar redshifts \edit1{\citep{kornei_properties_2012, martin_demographics_2012, bordoloi_dependence_2014,rubin_evidence_2014}}. If the absence of detectable outflows is an orientation effect, we would expect these four galaxies to be disks observed roughly edge-on.

To assess this possibility, we turn to the catalog of structural measurements of CANDELS galaxies \citep{van_der_wel_structural_2012}, from which we find that these four objects have axis ratios between 0.43 and 0.79, indistinguishable from the rest of the sample. 
We also find that these four galaxies are compact, all with effective radii below the sample median of 2.6 kpc; however, their SFRs are not particularly high, with all but one below the sample median, meaning that the high SFR surface densities are due more to small sizes than high SFRs. The compact nature of these objects limits our ability to derive constraints on their geometry and orientation, particularly given the fact that the WFC3 F160W images used in the \citet{van_der_wel_structural_2012} catalog are a factor of $\sim 2$ lower in spatial resolution than the ACS F606W and F814W images used by \citet{kornei_properties_2012} and \citet{bordoloi_dependence_2014}.

While this study is limited by both sample size and dynamic range, it demonstrates the novel combination of direct measurements of star formation at high redshift via \Ha emission line maps and direct measurements of outflows with deep absorption line spectroscopy. Our sample pushes the limits of current technologies:\ with existing ground-based facilities, full night integration times ($\sim 9$ hrs) are required to obtain sufficiently high S/N spectra of even bright galaxies at $z\gtrsim1$, while the time that would be required to obtain space-based \Ha maps of galaxies fainter than those in our sample is currently impractical. However, upcoming facilities will enable significant advances in both aspects of this study. The future extremely large telescopes (ELTs) will be able to obtain similarly high S/N rest-UV spectra of bright galaxies with a fraction of the time, and with longer exposures they will extend absorption line studies to fainter or more distant objects.

When the \textit{James Webb Space Telescope (JWST)} flies, it will conduct deep galaxy surveys at $1<z<6$, tracing star formation across $\sim 5$ Gyr of cosmic time and looking at earlier and more distant objects than \textit{HST} is able to see. Onboard, the Near-Infrared Imager and Slitless Spectrograph (NIRISS) will be analogous to Hubble's WFC3 camera. One of the four NIRISS observing modes enables wide field slitless spectroscopy over the entire field of view, using one or both of the telescope's grisms and blocking filters to isolate wavelength intervals between 0.8 and 2.2 $\mu m$. With \textit{JWST}'s increased collecting area, it will then be possible to map regions of star formation across a broader range of galaxies at $1\lesssim z \lesssim 2$. \textit{JWST} and the upcoming 30-m class telescopes will observe galaxies with lower masses and star formation rates, correlating outflows and star formation to measure galactic feedback across orders of magnitude in galaxy properties.

\acknowledgements
The authors thank \added{the referee for a thorough and constructive report, as well as} Gabriel Brammer for useful discussions and support with \textsc{Grizli}. N.Z.P. was supported by the University of Wisconsin-Milwaukee's Office of Undergraduate Research through the Support for Undergraduate Research Fellows (SURF) Award and Senior Excellence in Research Award (SERA). D.K.E. and N.Z.P. are supported by the US National Science Foundation (NSF) through the Faculty Early Career Development (CAREER) Program grant AST-1255591 and the Astronomy \& Astrophysics grant AST-1909198. C.L.M. is supported by NSF grant AST-1817125. The authors wish to recognize and acknowledge the very significant cultural role and reverence that the summit of Maunakea has always had within the indigenous Hawaiian community.  We are most fortunate to have the opportunity to conduct observations from this mountain.

\facilities{\textit{HST} (WFC3/G141), Keck:II (DEIMOS).}
\software{\textsc{Grizli} \citep{brammer_grizli_2019}, \texttt{astropy} \citep{the_astropy_collaboration_astropy_2013,the_astropy_collaboration_astropy_2018}, \texttt{AstroDrizzle} \citep{gonzaga_drizzlepac_2012}, DEEP2 Reduction Pipeline \citep{newman_deep2_2013,cooper_spec2d_2012}.}

\appendix
\section{Derivation of the Aperture of Highest Signal-to-Noise}
\label{app:sn}

We compute the optimal apertures that maximize the signal-to-noise ratio for background-limited sources with Gaussian and exponential surface brightness profiles as referenced in Section \ref{sec:sfr_sd} (the optimal radius of a Gaussian profile is a useful quantity for aperture photometry). We assume a two-dimensional circular surface brightness profile with radial form denoted by $g(r)$,
\begin{equation} \label{eqn:profile}
g(r) = \begin{dcases} 
          \exp\left(-\frac{r}{h}\right) & \text{exponential profile} \\
          \exp\left(-\frac{r^2}{2\sigma^2}\right) & \text{Gaussian profile},
       \end{dcases}
\end{equation}
where $h$ and $\sigma$ are the scale length and standard deviation of the exponential and Gaussian distributions respectively. We omit numerical prefactors in Equation \ref{eqn:profile}, since they do not affect the final result.

The signal $S(R)$ received from the source is the total number of photons collected by the detector, or equivalently, the number of photoelectrons produced.  For a source with enclosed area $\mathcal{A}$, $S(R)$ is given by $\displaystyle \int_{\mathcal{A}} g(r)\, \mathrm{d}A$, so for our assumed profiles
\begin{equation}
    S(R) = 2\pi\int_0^R g(r)\, r\, \mathrm{d}r.
\end{equation}
The variance of the signal is $S(R)$, assuming Poisson statistics.

The signal-to-noise of the observation is then given by the CCD equation \citep{merline_realistic_1995}: 
\begin{equation}
    S/N=\frac{S(R)}{\sqrt{S(R)+n_{\rm{pix}}\left(1+\frac{n_{\rm{pix}}}{n_B}\right)\left(N_B+N_D+N_R^2+G^2\sigma_f^2\right)}}.
\end{equation}
The number of pixels inside the photometric and background apertures correspond to $n_{\rm{pix}}$ and $n_B$ respectively, while the final term in the denominator contains the per pixel background noise ($N_B$), dark current ($N_D$), read noise ($N_R$), gain in electrons/ADU ($G$), and A/D conversion error ($\sigma_f$). 

We assume that the photometric aperture is small compared to the aperture used to determine the background, thus $n_{\rm{pix}}/n_B\ll 1$. To simplify the expression, we combine the non-Poisson sources of noise into a single per-pixel noise $\sigma_b$. For a circular aperture of radius $R$, the background variance is then $\pi R^2 \sigma_b^2$, and the total variance is 
\begin{equation}
    \sigma^2_{\rm{tot}} = S(R)+\pi R^2 \sigma_b^2.
\end{equation}
This leads to a signal-to-noise ratio
\begin{equation}
    S/N = \frac{S(R)}{\sqrt{S(R)+\pi R^2 \sigma_b^2}} = \frac{\displaystyle 2\pi\int_0^R g(r)\, r\,\mathrm{d}r}{\sqrt{\displaystyle 2\pi\int_0^R g(r)\, r\,\mathrm{d}r+\pi R^2 \sigma_b^2}},
\end{equation}
which we seek to maximize. 

Under the assumption of background-dominated observations, the Poisson noise is negligible compared to the background $\sigma_b$, and the signal-to-noise ratio is then  
\begin{equation}
    S/N \approx \frac{S(R)}{\sqrt{\pi R^2 \sigma_b^2}} = \frac{\displaystyle 2\pi\int_0^R g(r)\, r\,\mathrm{d}r}{\sqrt{\pi R^2 \sigma_b^2}}.
\end{equation}
In order to maximize the S/N, we numerically solve $\displaystyle \frac{\partial (S/N)}{\partial R}=0$ with the assumption that $S(R) \ll \pi R^2 \sigma_b^2$ for both the exponential and Gaussian profiles. The resulting radii corresponding to the highest S/N apertures are then:
\begin{equation}
R_{\rm{max}} = \begin{dcases} 
          1.79h & \text{exponential profile} \\
          1.59\sigma & \text{Gaussian profile}.
       \end{dcases}
\end{equation}
Note that for the Gaussian profile, $R_{\rm{max}}=0.67\,$FWHM.

\end{document}